\documentclass[twocolumn]{aastex62}
\pdfoutput=1 
\usepackage{amsmath,amstext}
\usepackage[T1]{fontenc}
\usepackage{apjfonts} 
\usepackage[figure,figure*]{hypcap}
\usepackage{gensymb}
\usepackage{array}

\newcolumntype{H}{>{\setbox0=\hbox\bgroup}c<{\egroup}@{}}

\shorttitle{43 \& 86 GHz SiO Maser Line Ratios}
\shortauthors{M.C. Stroh et al.}

\begin{document}

\title{Quasi-Simultaneous 43 and 86 GHz SiO Maser Observations and Potential Bias in the BAaDE Survey Are Resolved}

\author{Michael C. Stroh}
\affiliation{Department of Physics \& Astronomy, \\The University of New Mexico, Albuquerque, NM 87131}
\author{Ylva M. Pihlstr{\"o}m}
\altaffiliation{Y.\ M.\ Pihlstr{\"o}m is also an Adjunct Astronomer at the National Radio\\ Astronomy Observatory.}
\affiliation{Department of Physics \& Astronomy, \\The University of New Mexico, Albuquerque, NM 87131}
\author{Lor\'ant O. Sjouwerman}
\affiliation{National Radio Astronomy Observatory, \\Array Operations Center, Socorro, NM 87801}
\author{Mark. J. Claussen}
\affiliation{National Radio Astronomy Observatory, \\Array Operations Center, Socorro, NM 87801}
\author{Mark. R. Morris}
\affiliation{Department of Physics \& Astronomy, \\The University of California, Los Angeles, CA 90095}
\author{Michael R. Rich}
\affiliation{Department of Physics \& Astronomy, \\The University of California, Los Angeles, CA 90095}

    \received{2018 March 5}

    \revised{2018 June 10}

    \accepted{2018 June 13}

    \published{2018 August 2}

    \submitjournal{The Astrophysical Journal}

\begin{abstract}
We observed the 43 GHz $v$=1, 2 \& 3 and 86 GHz $v$=1 SiO maser transitions quasi-simultaneously for a Mira-variable dominated sample of over 80 sources from the Bulge Asymmetries and Dynamical Evolution (BAaDE) project, using ATCA, and statistically compared the relative line strengths. On average, the 43 GHz $v$=1 line is brighter than the 86 GHz $v$=1 line by a factor of 1.36$^{+0.15}_{-0.14}$. As a result, an 86 GHz $v$=1 observed sample can be observed to 85.9\% of the distance of a 43 GHz $v$=1 observed sample using the same sensitivity.  We discuss what impact this may have on the BAaDE Galactic plane survey using the VLA and ALMA. Despite fewer $v$=3 detections, specific trends are discerned or strengthened when the 43 GHz $v$=3 line is detected. In particular the 43 GHz and 86 GHz $v$=1 lines are on average equal for sources with no detectable 43 GHz $v$=3 emission, but the 43 GHz $v$=1 line strength is on average about twice as bright as the 86 GHz $v$=1 line for sources with detectable 43 GHz $v$=3 emission. Some weak correlations are found between line strengths and Midcourse Space Experiment (MSX) flux densities and colors, which are tightened when considering only sources with detectable 43 GHz $v$=3 emission. We discuss these trends in the context of a radiative pumping model to highlight how the 43 GHz $v$=3 line, when coupled with the $v$=1 and $v$=2 lines, can further our understanding of variable conditions like density in the circumstellar envelopes.
\end{abstract}


\section{Introduction}
The Bulge Asymmetries and Dynamical Evolution (BAaDE) project aims to explore the dynamical structures of the inner Galaxy and Galactic Bulge, using SiO maser lines from red giant stars. By constructing a sample of stellar point-mass probes, models of the gravitational potential can be tested using a final sample that is expected to provide about 20,000 line-of-sight velocities and positions. The SiO maser transitions occur at radio frequencies where extinction is negligible, thus allowing a dense sampling of line-of-sight velocities in the most crowded and optically obscured regions of the Milky Way. For a more detailed description of the BAaDE survey, see \citet{in...prep}.

While using both the Karl G.\ Jansky Very Large Array (VLA) and Atacama Large Millimeter/submillimeter Array (ALMA) ensures a full coverage of the Galactic plane, their different receiver availabilities (43 GHz at the VLA and 86 GHz at ALMA) also dictates that one part of the sample be observed in the 43 GHz $J$=1--0 SiO transitions (VLA), and the other part in the 86 GHz $J$=2--1 transitions (ALMA). A fundamental assumption for BAaDE is thus that stars emitting 43 GHz SiO maser emission also harbor 86 GHz masers, and vice versa. This appears to be a commonly accepted fact, supported, for example, by the work of \citet{2004PASJ...56...45S} who noted that out of 39 sources displaying 86 GHz SiO maser emission, 38 also produced 43 GHz masers. What is less clear, however, is whether there is a statistically significant difference in the integrated flux densities of 43 GHz versus 86 GHz masers. Such a difference could have an impact on the analysis of the BAaDE sample, as the VLA and ALMA samples are in principle observed to the same noise levels and two different Malmquist biases would have to be considered when combining the two samples to draw statistical conclusions. This is especially important, as the 86 GHz ALMA sample, which utilizes the possibly fainter transition of the masers (see below), covers Galactic longitudes of $-110\degree<l<0\degree$, and therefore contains the region of the bar that is furthest from us.

Observations of both 43 GHz and 86 GHz SiO maser transitions indicate that the flux-density ratio may depend on the stellar type, or circumstellar shell thickness.  The shell thickness can be inferred from infrared colors, as the thicker envelopes produce redder colors.  In the optically-thick case, the central star may be completely obscured in the optical regime, while it is observable for the optically thin shells. For Asymptotic Giant Branch (AGB) stars, this has been outlined by \citet{1988A&A...194..125V} using IRAS color-color diagrams, where thick-shell OH/IR stars are found in the redder regions IIIb and IV, and thin-shell Mira variables, with typical pulsation periods longer than 100 days and IR amplitude variations greater than one magnitude, in the bluer regions II and IIIa.

\citet{1993A&A...280..551N} observed 40 OH/IR stars and noted that the integrated flux densities of the 43 GHz $v$=1 lines were brighter than the 86 GHz $v$=1 lines (which were observed on average two weeks later), and that the difference was greater for stars with thicker envelopes compared to those with thinner envelopes. Similarly, \citet{2016JKAS...49..261K} report on the 43 GHz v=1 line being brighter than the 86 GHz v=1 in two additional OH/IR stars.

The BAaDE sample was chosen from a sample of stars likely to be dominated by Miras and hence possess much thinner envelopes than the OH/IR stars. This sample, by an extension of the results of \citet{1993A&A...280..551N}, might imply a smaller difference between the 43 GHz and 86 GHz line flux densities, and perhaps even a reversal in which the 86 GHz line would be brighter. This would be consistent with a weak trend reported by \citet{1998A&A...329..219P}, where 5 out of 9 sources were brighter in the 86 GHz v=1 line compared to the 43 GHz v1 (the majority of which were Miras). Further, numerical simulations investigating collisionally pumped maser time variability in Miras suggests that, averaged over the stellar cycle, the $v$=1 maser emission is expected to be brighter at 86 GHz than at 43 GHz \citep{2002A&A...386..256H}. However, this collisionally pumped model does not include the effects of line overlap at the pumping frequencies, which can produce quite different relative line strengths in radiatively pumped models \citep[e.g.,][]{2014A&A...565A.127D}. For the BAaDE sample, we have no information about the stellar phase for individual targets, thus averaging information over a large sample of stars will be considered to effectively correspond to a full-sample cycle average.

To assess the robustness of the assumption that the strengths between the 43 GHz and 86 GHz SiO maser emission are on average equal, we use quasi-simultaneous observations of the 43 and 86 GHz SiO lines in a sample of BAaDE sources.

In section 2 we describe the selected sample, the observations and data calibration.
In sections 3 and 4 we share our results and discuss the significance between various line strengths starting with a comparison between the 43 and 86 GHz $v$=1 lines.

\section{Observations and Data Analysis}
The Australia Telescope Compact Array (ATCA) allows the 43 GHz and 86 GHz SiO transitions to be observed quasi-simultaneously with sufficient signal-to-noise within a limited time. We selected 86 targets\footnote{see the Appendix for source coordinates.} detected in SiO maser emission in the pilot BAaDE survey. The sample resides in the longitude range $-16\degree<l<8\degree$ in order to contain sources previously detected by the VLA and ALMA programs that could be observed on the same day. 34 of those were detected with the VLA in the 43 GHz lines, and 52 with ALMA in the 86 GHz lines. Given that the 86 GHz system at ATCA requires much longer integration times compared to the 43 GHz to reach a similar rms noise, the targets selected were among the brightest masers in the BAaDE survey which may bias this ATCA sample towards closer or inherently brighter sources.

The Compact Array Broadband Backend (CABB) allows several SiO transitions to be observed at the same time. With the CFB64M-32k mode and the full set of 16 zoom bands, three transitions at 43 GHz and three transitions at 86 GHz were observed (Table \ref{spw_velocities}). Each spectral channel is 31.25 kHz wide, providing a velocity resolution of 0.22 km/s (0.11 km/s) at 43 GHz (86 GHz). For each source, the center frequency of each group of zoom bands was set to the sky frequency, using the Local Standard of Rest (LSR) velocity of the lines previously detected with ALMA or the VLA.

\begin{deluxetable}{lcrrrr}[t]
\tablecaption{ATCA observations spectral line coverage\label{spw_velocities}}
\tabletypesize{\scriptsize}
\tablehead{
\colhead{} & \colhead{Rest Frequency} & \colhead{Zoom} & \colhead{Bandwidth} & \colhead{Bandwidth} & \colhead{Number of}\\
\colhead{Line Name} & \colhead{(MHz)} & \colhead{Bands} &  \colhead{(MHz)} & \colhead{(km~s$^{-1}$)} & \colhead{Detections}
}
\startdata
43 GHz $v$=1 & 43~122.09 & 3 & 121 & 847 & 81\\
43 GHz $v$=2 & 42~820.50 & 7 & 249 & 1749 & 81\\
43 GHz $v$=3 & 42~519.28 & 6 & 217 & 1536 & 31\\
\hline
86 GHz $v$=0 & 86~846.99 & 4 & 153 & 532 & 3\\
86 GHz $v$=1 & 86~243.37 & 5 & 185 & 646 & 66\\
86 GHz $v$=2 & 85~640.45 & 7 & 249 & 875 & 4\\
\enddata
\end{deluxetable}

Our 86 sources were observed over seven dates in 2016: July 25th, 27th, 28th and 29th, and September 3rd, 4th and 8th.
Each source was observed first at 43 GHz followed by scans at 86 GHz on the same day for which the observing dates for each source are listed in Table \ref{maser_summary_table}.
The time between the 43 GHz and 86 GHz scans varied between 7 and 120 minutes with an average of 25 minutes; therefore, since stars harboring SiO emission may have stellar periods of $\sim 80-1000$ days, we do not expect the stellar variability to significantly impact our results.
Each source was observed using a single scan with integration time per source averaged three minutes at 43 GHz and six minutes at 86 GHz. Observations used five of the six antennas in configurations H75 and H168 for our July and September observations, respectively. Typical rms noise of $\sim 0.07$ ($\sim 0.2$) mJy / beam / channel at 43 (86) GHz were measured. 

Each spectral window was calibrated separately using MIRIAD \citep{1995ASPC...77..433S}.
Uranus was used for flux density calibration on four of our observing dates (July 28th, July 29th, September 3rd and September 8th) and the flux density calibration was bootstrapped using 3C 279, which was observed on all observing dates. 3C 279 also served as a bandpass calibrator along with QSO B1921-293.
Phase calibration was performed using QSO B1830-211 and 1729-37.
For the detected maser lines, self-calibration was applied and the data were averaged over 1 $\textnormal{km~s}^{-1}$ for final line detections.

A line is considered a detection if the velocity-integrated flux density over 1 $\textnormal{km~s}^{-1}$ is at least five times the rms noise. For each detected line, a final velocity-integrated flux density is calculated by integrating over a 3 $\textnormal{km~s}^{-1}$ range centered on the peak.

\section{Results}
From the 86 sources observed, SiO 43 GHz $v$=1, 2 \& 3 emission was detected in 81, 81 and 31 sources respectively, and SiO 86 GHz $v$=0, 1 \& 2 emission was detected in 3, 66 and 4 sources, respectively, at the 5$\sigma$ level. The velocity-integrated flux density for each detected line is presented in Table \ref{maser_summary_table}.

The BAaDE VLA observations of our sample sources were carried out on March 15th, 2013, over three years prior to our ATCA observations. The BAaDE observations of the ALMA sources were carried out between December 27th, 2016 and January 18th, 2016, thus many of the ALMA sources have not experienced a full stellar cycle since the original BAaDE observations and may not represent a random sampling of stellar phases since sources were among the brightest in the BAaDE survey.

\subsection{Detection Rates}
The 43 GHz $v$=1 line was detected in 81 of the 86 sources in our sample. Of the five 43 GHz $v$=1 non-detections, four of the upper limits are below the previously detected flux density measured in the VLA observations, likely manifesting a large variability of the SiO maser causing it to fall below our detection limit.  Observations of SiO maser variability have revealed factors of 20-50 over a stellar cycle, and that some features completely disappear between cycles \citep[e.g.,][]{1986A&A...159..166G, 2005PASJ...57..341K, 2006ApJ...638L..41M, 2006ApJ...649..406M}. The fifth non-detection was in a source from the ALMA sample, thus no prior 43 GHz information was available.

The 43 GHz $v$=2 line was also detected in 81 of the 86 sources. Out of the five non-detections, two have neither 43 GHz $v$=1 nor $v$=2 ATCA detections. The VLA flux densities of two non-detections are below the ATCA rms noise level, while three should have been detectable, implying varying maser flux densities. 

Since the 43 GHz $v$=3 transition needs additional energy to be pumped compared to the $v$=1 and $v$=2 levels, fewer $v$=3 detections are expected. This is consistent with our 31 detections in the 43 GHz $v$=3 line and we note that our 36\% detection rate of this line is close to the 36.7\% detection rate of this line by \citet*{2007ApJ...669..446N}, although their sample included sources from most regions in the IRAS two-color diagram. Additionally \citet*{1996AJ....111.1987C} suggested that this line is most often observed when the source is at its brightest infrared phase.

The 86 GHz $v$=1 line was detected in 66 sources. Out of the 20 sources where the 86 GHz $v$=1 line was not detected, four are from our ALMA sample and were observed during poor weather conditions on the September observing dates, which greatly increased the rms noise. Three of these upper limits are consistent with previous ALMA detections. The fourth of these non-detections could indicate source variability. The remaining 16 non-detections are from the VLA sample where we have no previous 86 GHz information, and 11 of these were also observed during poor weather conditions.

The SiO 86 GHz $v$=0 lines were only detected in three sources above the 5$\sigma$ level, but this is a thermal line and requires much longer exposure times to detect. We have therefore not used any of these data in our study here, but for completeness we list the results in Table \ref{maser_summary_table}. Similarly for the $v$=2 line discussed next.

The SiO 86 GHz $v$=2 line is only detected in four of our sources. The anomalous weakness in the strength of this line has been observed since the 1970s and an explanation for it was first proposed by \citet{1981ApJ...247L..81O} who noted the close proximity between the 8$\mu$m transitions of $J$=0--1 $v$=1--2 SiO \citep{1974ApJ...191L..37G} and $v$=0--1 $J$=12$_{7,5}$--11$_{6,6}$ para-H$_2$O transition \citep{1952_benedict}. This overlap and its effect on the pumping models has been studied by \citet{1996A&A...314..883B} and \citet{2014A&A...565A.127D}.

In summary, we re-detect the majority of sources previously detected with VLA and ALMA, as expected based on their VLA and ALMA flux densities and the ATCA sensitivity. Most of the non-detections can be explained with a poor signal-to-noise ratio. However, this does not explain all non-detections; some of the sources likely varied downward since the previous observations and thus fall below the detection limit.

\subsection{Line Ratios}
To compare the relative line strengths, Table \ref{maser_summary_table} includes line ratios for a 3 km~s$^{-1}$ velocity interval integrated around the peak channel. Although integrated flux densities, $I$, are calculated using 3 km~s$^{-1}$ velocity intervals, similar results for line ratios are found in our data using 1 or 5 km~s$^{-1}$ intervals. In order to minimize the effects of bad weather conditions on the results, for the few sources observed on multiple dates, only data collected on dates with the lowest rms noise are considered while ensuring that data for individual sources were collected on the same day.

For defining our line ratios, in order to remain consistent with previous comparisons for line ratios, we have adopted the standard of placing the higher frequency line in the denominator, thus referring to $\frac{I(\textnormal{43~GHz}~v=1)}{I(\textnormal{86~GHz}~v=1) }$, $\frac{I(\textnormal{43~GHz}~v=2)}{I(\textnormal{43~GHz}~v=1) }$ and $\frac{I(\textnormal{43~GHz}~v=3)}{I(\textnormal{43~GHz}~v=1) }$.

While the simple average of line ratios was sufficient for \citet{1993A&A...280..551N} to differentiate between their thin- and thick-shell sources, it is not appropriate when the ratio of line strengths approaches unity, since the distribution of the ratios becomes skewed. We will instead be discussing the distributions of logarithmic line ratios throughout this paper which will be unbiased to the choice of numerator and denominator in the ratio, and in particular be interested in the centers and standard deviations of these distributions. For example \citet{1993A&A...280..551N} found $\langle \frac{I(\textnormal{43~GHz}~v=1)}{I(\textnormal{86~GHz}~v=1) }\rangle \approx 2$ for their thinner envelope population which is similar to the simple average of our population, i.e. $\langle \frac{I(\textnormal{43~GHz}~v=1)}{I(\textnormal{86~GHz}~v=1) }\rangle=1.97 \pm 0.24$. However, due to the large range of values for the integrated intensities in our sample, we find $\langle \frac{I(\textnormal{86~GHz}~v=1)}{I(\textnormal{43~GHz}~v=1) }\rangle=1.06 \pm 0.12$ (i.e.,\ the average of the reciprocal of line ratios is also larger than one). Considering instead the distribution of the logarithmic ratios, we find that on average $I(\textnormal{43~GHz}~v=1) > I(\textnormal{86~GHz}~v=1)$ using either $\log\Big[\frac{ I(\textnormal{43~GHz}~v=1) }{ I(\textnormal{86~GHz}~v=1) }\Big]$ or $\log\Big[\frac{ I(\textnormal{86~GHz}~v=1) }{ I(\textnormal{43~GHz}~v=1) }\Big]$.

\section{Discussion}

\subsection{Comparing SiO 43 GHz and 86 GHz $v$=1 Lines}

\begin{figure*}[t]
\figurenum{1}
\includegraphics[scale=0.47]{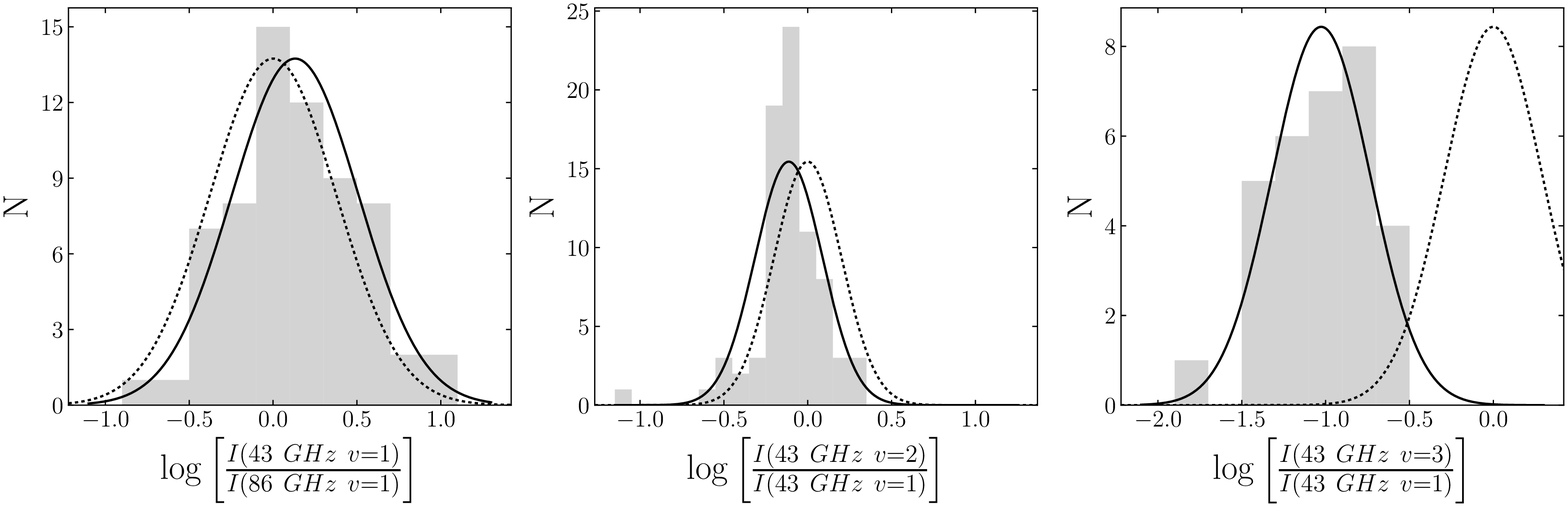}
\caption{From left to right, histograms of $\log\Big[\frac{ I(\textnormal{43~GHz}~v=1) }{ I(\textnormal{86~GHz}~v=1) }\Big]$, $\log\Big[\frac{ I(\textnormal{43~GHz}~v=2) }{ I(\textnormal{43~GHz}~v=1) } \Big]$, and $\log\Big[ \frac{ I(\textnormal{43~GHz}~v=3) }{ I(\textnormal{43~GHz}~v=1) }\Big]$. Left and right adopt a bin size of 0.2 and the center uses a bin size of 0.1. Left: The center and standard deviations of the distribution are $0.13 \pm 0.05$ and $0.38 \pm 0.03$ respectively. The center of the distribution implies that the 43 GHz $v$=1 line on average is 1.36$^{+0.15}_{-0.14}$ times brighter than the 86 GHz $v$=1 line. Center: The center and standard deviations of distribution (solid line) are $-0.11 \pm 0.02$ and $0.20 \pm 0.02$ respectively, indicating that the 43 GHz $v$=2 line strength tends to be a factor of $0.77\pm 0.04$ as strong as the $v$=1 line. Right: The center and standard deviations of the distribution (solid line) are $-1.03 \pm 0.05$ and $0.29 \pm 0.03$ respectively, implying that the 43 GHz $v$=3 line, when detected, is on average an order of magnitude weaker than the 43 GHz $v$=1 line. Dotted lines represent Gaussian distributions for the same standard deviations and heights, but placed where the lines would be if lines were of equal strength on average (i.e., with center at 0).}
\label{fig_consolidated_histogram}
\end{figure*}

This study aims to determine whether BAaDE selected sources tend to be brighter in the 86 GHz $v$=1 transition than the 43 GHz $v$=1 transition or vice versa, and whether this needs to be accounted for in analyzing the survey. 65 sources from our quasi-simultaneous observations displayed both 43 and 86 GHz $v$=1 emission lines.
Figure \ref{fig_consolidated_histogram}, left, displays the distribution of line ratios, plotted as number versus log(ratio).

The Anderson-Darling non-parametric test is used on the distribution of ratios in Fig.\ \ref{fig_consolidated_histogram}, left, with a null hypothsis that the distribution is Gaussian. The Anderson-Darling test gives a p-value of 0.9, thus we cannot reject the null hypothesis, and we assume that the distribution of ratios is Gaussian. The \textit{fitdistr} function \citep[residing in the \textit{MASS} \textit{R} package, see][]{ISBN0-387-95457-0} provides maximum-likelihood estimation of univariate distributions using a number of probability distributions. Assuming the distribution of ratios is Gaussian, the maximum-likelihood regression shows that the center and standard deviation of the distribution are $0.13 \pm 0.04$ and $0.38 \pm 0.03$, respectively. The center of the distribution of ratios being shifted from zero, i.e. the logarithm of unity, is evidence that the 43 GHz line is on average stronger than the 86 GHz $v$=1 line.
The distribution suggests a 64\% chance that for a randomly chosen source, the $v$=1 line will be stronger at 43 than at 86 GHz. To further confirm the significance of the difference in integrated flux densities for the $v$=1 lines, the Wilcoxon signed-rank test can also be used.

The Wilcoxon signed-rank test \citep{10.2307/3001968} is a non-parametric paired-difference test and can be used as an alternative to the paired Student's t-test for arbitrary distributions. In this case, the test assumes that on average the integrated flux densities of the $v$=1 lines for each source are equal, and the test is appropriate for comparing the integrated flux densities since it does not assume that either the 43 GHz $v$=1 or 86 GHz $v$=1 integrated flux densities are drawn from a particular distribution. The test gives a p-value of 0.0018 with the alternative hypothesis of the 43 GHz $v$=1 line being stronger than the 86 GHz $v$=1 line. Since this p-value is less than 0.01 (corresponding to a 99\% confidence level), the null hypothesis is rejected and the alternative hypothesis is adopted. Consequently, there is evidence that the detected line strengths are statistically different on average, but the large scatter in the ratios leads to the full-width half maximum of the distribution in Fig.\ \ref{fig_consolidated_histogram}, left, representing nearly an order of magnitude range in the 43 to 86 GHz $v$=1 line ratios.

\begin{figure}
\figurenum{2}
\includegraphics[scale=0.67]{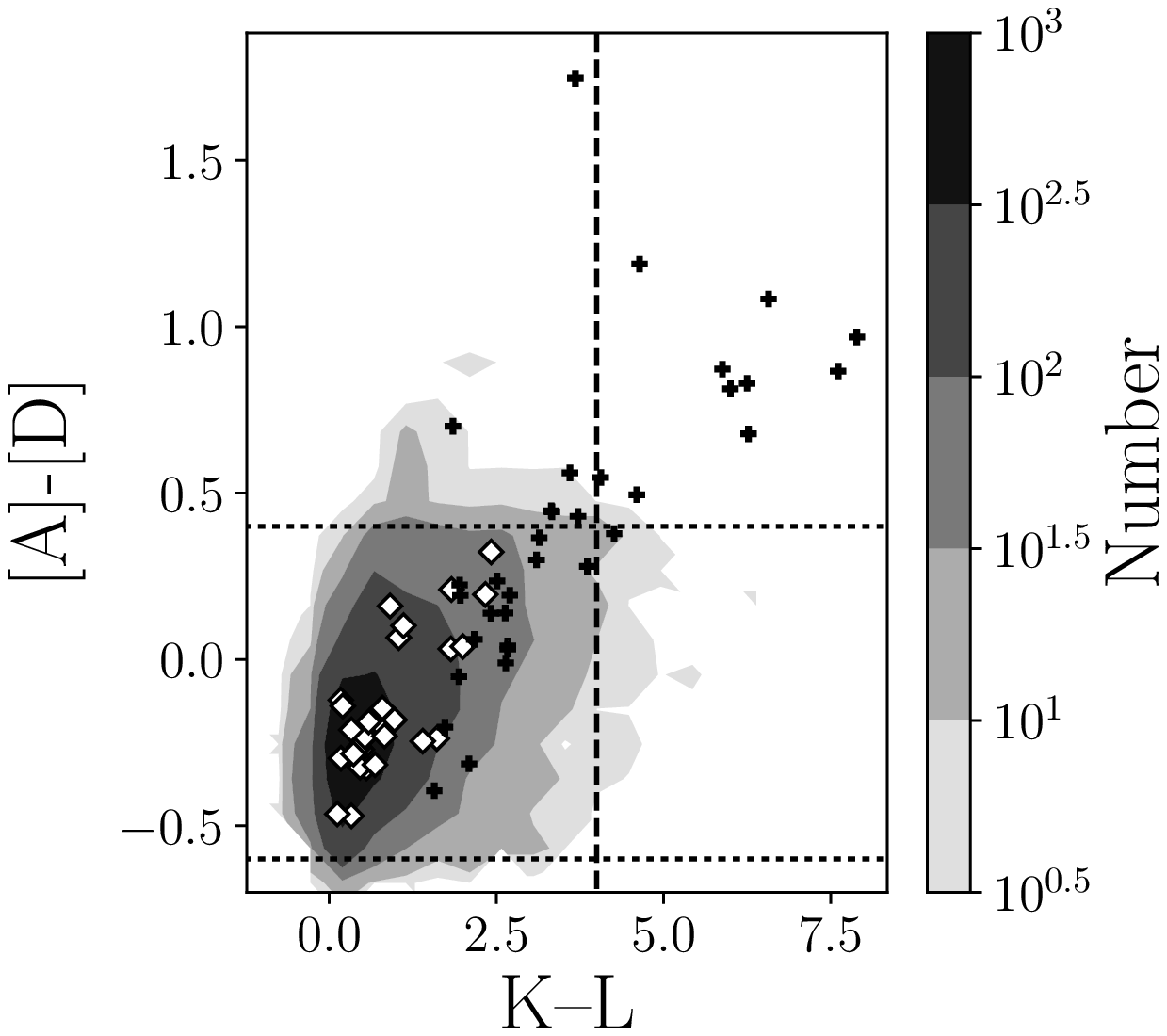}
\caption{Two-color diagram of the $\approx$9000 BAaDE sources which have both 2MASS K and WISE band 1 measurements used as a proxy for K--L (filled contours). The contours are calculated using a grid of 20 bins spanning $-1.2 \lessapprox \textnormal{K}-\textnormal{L} \lessapprox 7.8$ and 20 bins along $-1 \lessapprox [A]-[D] \lessapprox 1$ for a total of 400 bins.  The subset of BAaDE sources used in this study represented by white diamonds and targets from \citet{1993A&A...280..551N} are represented as black crosses. The horizontal lines represent the primary color selection (i.e.\ $-0.6 \le [A]-[D] \le 0.4$) used to choose sources in region \textit{iiia} of the Midcourse Space Experiment (MSX) two-color diagram \citep{2009ApJ...705.1554S}. The vertical line at K--L = 4 is where Nyman et al. suggested a change from thinner to thicker envelopes in OH/IR stars. The majority of the sources from \citet{1998A&A...329..219P} have K--L < 4 using the same uncorrected proxy for K--L that we use for the BAaDE sample.}
\label{fig_nyman_baade_comparison}
\end{figure}

As mentioned in Sect.\ 1, \citet{1993A&A...280..551N} investigated the SiO maser line ratios in a sample of OH/IR stars. They found that thicker shells have a brighter 43 GHz $v$=1 transition compared to the 86 GHz $v$=1, using K (2.19$\mu$m)--L (3.79$\mu$m) colors to define the relative shell thickness. The BAaDE sample is selected to have thinner shells than the OH/IR stars, with a large fraction likely being Miras. The full BAaDE sample \citep{in...prep} is comprised of over 28,000 sources from color selections of [A]--[D], [A]--[E] and [C]--[E] corresponding to region \textit{iiia} of the MSX two-color diagram \citep{2009ApJ...705.1554S}, where MSX A, C, D and E bands are centered on wavelengths of 8.28, 12.13, 14.65 and 21.34$\mu$m, respectively. Approximately 9,000 BAaDE sources have both associated 2MASS K band (2.19$\mu$m) and WISE band 1 (3.8$\mu$m) matches, and a close proxy to the K--L measurements provided by Nyman et al.\ can be formed. In Fig.\ \ref{fig_nyman_baade_comparison} the distribution of the 9,000 BAaDE targets, the BAaDE ATCA sample, as well as the Nyman OH/IR sample is plotted in a color-color diagram ([A]--[D] versus K--L). The BAaDE sample has a mean K--L color of 1.1, with only 1\% of the sources having K--L>4. Thus, the BAaDE sample fits the thin envelope criteria of Nyman et al., and the BAaDE survey may serve as a thin shell extension to the thicker shell OH/IR star sample studied by Nyman et al. The BAaDE subsample distribution in Fig.\ \ref{fig_nyman_baade_comparison} further illustrates that this sample is representative of the [A]--[D] range probed by the full BAaDE survey.

Collisionally pumped modeling by \citet{2002A&A...386..256H} suggested that averaged over a stellar cycle, the $v$=1 line should be stronger at 86 GHz than at 43 GHz in Miras. Since our observations may occur at random times during the sample sources' stellar cycles, if the 86 GHz $v$=1 line were on average brighter over the stellar cycle, this could be evident in the ATCA sample. However, our results demonstrate (see Sect.\ 4.5) that there is no apparent transition into a regime where the 86 GHz $v$=1 line is brighter than the 43 GHz $v$=1 line when moving from thicker shell OH/IR stars into thinner shell Mira-like objects. Note that the collisionally pumped models by \citet{2002A&A...386..256H} did not include pumping line overlap. Instead, the radiatively pumped \citet{2014A&A...565A.127D} models, where the SiO--H$_2$O line overlap is taken into account, suggest that while there exists a regime of lower shell densities for which the 86 GHz $v$=1 should be brighter, for higher number densities the 43 GHz $v$=1 line is expected to be the stronger line. The \citet{2014A&A...565A.127D} model further predicts a large range of relative line ratios, depending on the shell density. Such a large spread of possible values could perhaps explain some of the previous results \citep[e.g.,][]{1993A&A...280..551N,1998A&A...329..219P} which hinted at a brighter 86 GHz $v$=1 line. The BAaDE ATCA sample is larger than in previous studies, which may improve the sampling of the line ratios especially given the large spread.

\subsubsection{Impact on the BAaDE Survey}
Since on average, the $v$=1 line is weaker at 86 GHz than at 43 GHz, we need to determine how this will affect the volume that the BAaDE survey will be able to probe. If both the ALMA and VLA BAaDE samples are observed to approximately the same noise level, the center of the distribution in Fig.\ \ref{fig_consolidated_histogram}, left, (i.e. $\log\Big[\frac{ I(\textnormal{43~GHz}~v=1) }{ I(\textnormal{86~GHz}~v=1) }\Big] = 0.13 \pm 0.05$) gives an intrinsic relative source brightnesses of the two $v$=1 lines are related by $\frac{I(\textnormal{43~GHz}~v=1)}{I(\textnormal{86~GHz}~v=1)} = 1.36^{+0.15}_{-0.14}$, then on average 86 GHz sources can only be observed to 85.9\% of the distance that the same source could be observed at using the 43 GHz transitions.

In the context of the Galactic long bar with a radius of 4.0 kpc \citep{2005ApJ...630L.149B, 2007A&A...465..825C}, a distance to the Galactic Center of 8.34 kpc \citep{2014ApJ...783..130R} and that the bar lies at an approximately 30$\degree$ angle to our line of sight \citep{2015MNRAS.450.4050W}, the far side of the long bar is approximately 12 kpc away. Consequently a source located at the far side of the long bar detected at 43 GHz at our sensitivity limit, would have to be moved 1.8 kpc closer to the Galactic Center in order to be detected in its 86 GHz line while remaining on the bar. Thus, this bias could have an appreciable effect on the sampled BAaDE population. However, the large spread in the line ratios implies that 86 GHz $v$=1 emission is brighter than the 43 GHz $v$=1 line in nearly 40\% of the sources, thus we expect to be able to still have a large number of sources in our sample that can be observed in the furthest regions of the Galactic Plane.

We note that since our sample contains many of the brightest stars in the full BAaDE sample, these sources may be closer and thus we may be not sampling among the full age and metallicities of stars in the the full BAaDE survey. Therefore, studies that can probe this line ratio throughout the Milky Way disk are worthwhile.

\subsection{Comparing SiO 43 GHz $v$=1 and $v$=2 Lines}

To examine the relationship between the 43 GHz $v$=1 and $v$=2 lines, the integrated flux densities of the SiO 43 GHz $v$=1 and $v$=2 lines were compared using the 78 sources\footnote{While 43 GHz $v$=1 and 2 were individually detected in 81 sources, 43 GHz $v$=1 was detected in three sources where no $v$=2 emission was detected. Similarly 43 GHz $v$=2 emission was detected in three sources where no $v$=1 emission was detected. See Table \ref{maser_summary_table} for more information.} where both lines were detected. Figure \ref{fig_consolidated_histogram}, center, shows the line ratio distribution for $\log\Big[\frac{ I(\textnormal{43~GHz}~v=2) }{ I(\textnormal{43~GHz}~v=1) }\Big]$. The center of the maximum-likelihood regression to a Gaussian distribution, using \textit{fitdistr}, is at $-0.11 \pm 0.02$ with a standard deviation of $0.20 \pm 0.02$ which suggests a 71\% chance that for any BAaDE source, the 43 GHz $v$=2 line will be weaker than the 43 GHz $v$=1 line.

However, we reject the null hypothesis that the distribution is Gaussian at the 99\% confidence level using the Anderson-Darling test which gives a p-value $\sim10^{-4}$. The distribution is likely better described by a two component model with a narrow component to account for the strong, tightly coupled relation between the $v$=1 \& 2 lines, and a second component that accounts for the weaker broader distribution. Coupling line ratios to stellar variability, density, temperature, radiation environment, and mass loss rates may lead to a better understanding of the contributions that shape this distribution. For example, \citet{2016JKAS...49..261K}'s observations of post-AGB stars indicate that the $v$=2 line is stronger than the $v$=1 line for both integrated and peak line ratios. Thus the more evolved population may reside in the right hand tail of the distribution. We note that this distribution appears consistent with previous results \citep[e.g., see Fig. 4 of ][]{1996AJ....111.1987C}, but this dispersion is much smaller than that of the 43 GHz $v$=1 to 86 GHz $v$=1 line ratios in the previous subsection.

Assuming a null hypothesis that $I(\textnormal{43~GHz}~v=2) \ge I(\textnormal{43~GHz}~v=1)$, the Wilcoxon signed-rank test produces a p-value $\sim 10^{-7}$ with the alternative that $I(\textnormal{43~GHz}~v=1) > I(\textnormal{43~GHz}~v=2)$. Since this p-value is less than 0.01, at the 99\% confidence level we accept the alternative that at 43 GHz for sources in the BAaDE sample, the $v$=2 line is on average weaker than the $v$=1 line. Although both the histogram and Wilcoxon signed-rank tests indicate a clear difference in line strengths, the full VLA BAaDE sample will dwarf our ATCA sample size by orders of magnitude and more thoroughly test any trends observed here since the 43 GHz $v$=0, 1, 2 \& 3 lines, as well as three isotope lines, are all observed simultaneously in the BAaDE survey.

\subsection{Comparing SiO 43 GHz $v$=1 and $v$=3 Lines}
All 31 sources where the 43 GHz $v$=3 transition was detected had both 43 GHz $v$=1 and $v$=2 line detections. The Anderson-Darling test gives a p-value of 0.7 for the null hypothesis that the distribution in Fig.\ \ref{fig_consolidated_histogram}, right, is Gaussian. Thus with the high p-value, we accept that the distribution is Gaussian.

A maximum-likelihood regression to a Gaussian distribution using \textit{fitdistr} yields a center and standard deviation of $-1.03 \pm 0.05$ and $0.29 \pm 0.04$, respectively, for the $\log\Big[\frac{ I(\textnormal{43~GHz}~v=3) }{ I(\textnormal{43~GHz}~v=1) }\Big]$ distribution. When the 43 GHz $v$=3 line is detected, its line strength is on average an order of magnitude weaker than the 43 GHz $v$=1 and $v$=2 line strengths. The dispersions of the $\frac{I(\textnormal{43~GHz}~v=3)}{I(\textnormal{43~GHz}~v=1)}$ and $\frac{I(\textnormal{43~GHz}~v=1)}{I(\textnormal{86~GHz}~v=1)}$ logarithmic distributions are similar.


\begin{figure}
\figurenum{3}
\includegraphics[scale=0.6]{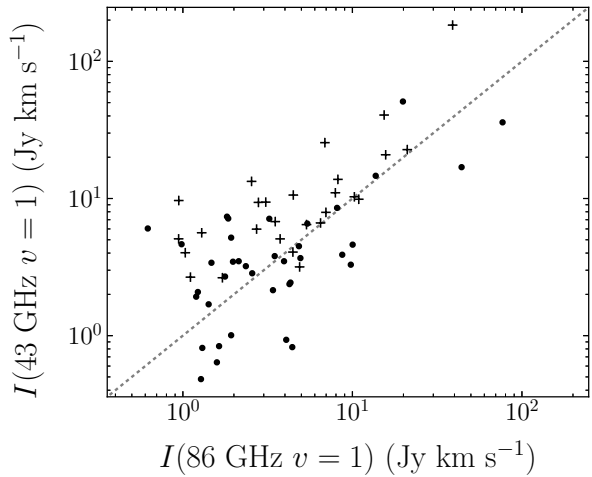}
\caption{Comparison between $I(\textnormal{43~GHz}~v=1)$ and $I(\textnormal{86~GHz}~v=1)$ for sources where both lines are detected. The dashed line indicates where lines are of equal strength. Crosses represent sources where the $v$=3 line was detected, and tend to be associated with higher relative 43 GHz $v$=1 emission.}
\label{fig_43_86_v1_scatter_with_v3}
\end{figure}

\begin{figure}
\figurenum{4}
\includegraphics[scale=0.57]{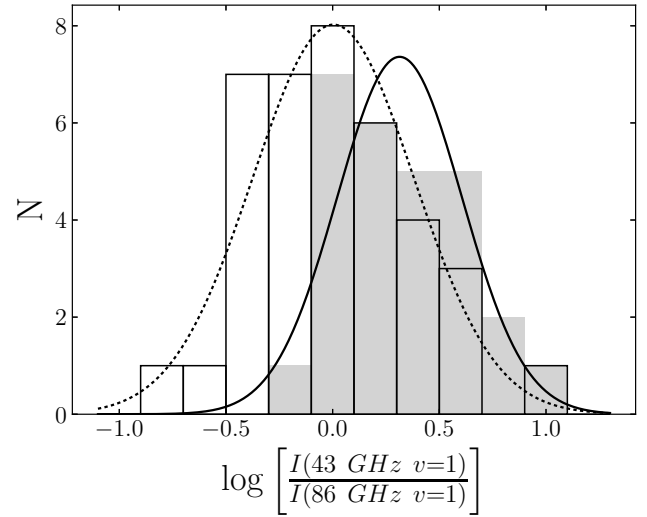}
\caption{Histogram of $\log\Big[\frac{ I(\textnormal{43~GHz}~v=1) }{ I(\textnormal{86~GHz}~v=1) }\Big]$ distributions for the populations where the 43 GHz $v$=3 line was detected in gray, and where the line was not detected are outlined. For sources with 43 GHz $v$=3 detected lines, the center and standard deviations of the distribution (solid line) are $0.31\pm0.06$ and $0.29\pm0.04$, respectively. For sources without 43 GHz $v$=3 detected emission, the center and standard deviations of the distribution (dotted line) are $0.00\pm0.06$ and $0.38\pm0.04$, respectively. Thus on average, the sources with no 43 GHz $v$=3 detections, the emission from 43 and 86 GHz $v$=1 maser lines are equally strong, but for sources with 43 GHz $v$=3 line detections, the $v$=1 line is on average brighter at 43 GHz than 86 GHz (e.g. $I(\textnormal{43~GHz}~v=1)$ is on average $2.0\pm 0.3$ times as luminous as $I(\textnormal{86~GHz}~v=1)$).}
\label{fig_43_86_with_v3_histogram}
\end{figure}

\subsection{Trends for the 43 GHz $v$=3 Line Detected Sources}
The presence or absence of maser emission from higher energy vibrational levels, such as the $v$=3 lines, is useful for investigating pumping scenarios when combined with the $v$=1 and 2 transitions. The relative line intensities should reflect the combined pumping conditions due to collisions and/or radiation, both of which vary as a function of the phase of the stellar cycle \citep{2002A&A...386..256H, 2014A&A...565A.127D}. For example, \citet{1996AJ....111.1987C} noticed that the 43 GHz $v$=3 lines were present when the star was near the maximum IR luminosity phase of the stellar pulsation cycle, which would likely also be associated with a specific density phase.

Figure \ref{fig_43_86_v1_scatter_with_v3} shows that $I(\textnormal{43~GHz}~v=1)$ is almost always larger than $I(\textnormal{86~GHz}~v=1)$ for sources where the 43 GHz $v$=3 line is detected in the BAaDE ATCA sample. Figure \ref{fig_43_86_with_v3_histogram} shows the distribution of 43 to 86 GHz $v$=1 line ratios for sources with and without 43 GHz $v$=3 lines respectively. The Anderson-Darling test cannot reject the null hypotheses that both $\log\Big[\frac{ I(\textnormal{43~GHz}~v=1) }{ I(\textnormal{86~GHz}~v=1) }\Big]$ distributions are Gaussian at the 99\% confidence level, since it gives p-values 0.4 and 0.7 for the distributions of sources with and without $v$=3 line detections, respectively. A maximum-likelihood regression to a Gaussian distribution using \textit{fitdistr} shows that the center and standard deviation for the distribution without $v$=3 detections to be $0.00\pm0.06$ and $0.38\pm0.04$, respectively, thus this is consistent with the two $v$=1 lines being of equal strength. A similar regression for the distribution of sources with $v$=3 detections has center and standard deviations of $0.31\pm0.06$ with standard deviation $0.29\pm0.04$.


\begin{figure}
\figurenum{5}
\includegraphics[scale=0.6]{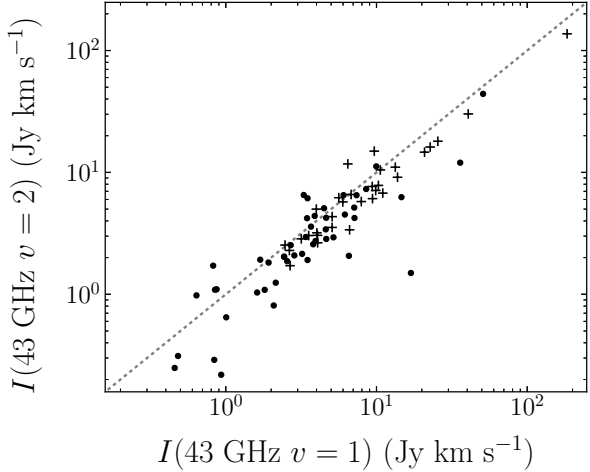}
\caption{Comparison between $I(\textnormal{43~GHz}~v=2)$ and $I(\textnormal{43~GHz}~v=1)$. Crosses represent sources where the $v$=3 line was also detected. The dashed line indicates where lines are of equal strength, showing most sources have a relatively brighter $v$=1 line and that $v$=3 detected sources tend to have a smaller dispersion from the overall trend.}
\label{fig_43_v2_v1_scatter_with_v3}
\end{figure}

\begin{figure}
\figurenum{6}
\includegraphics[scale=0.6]{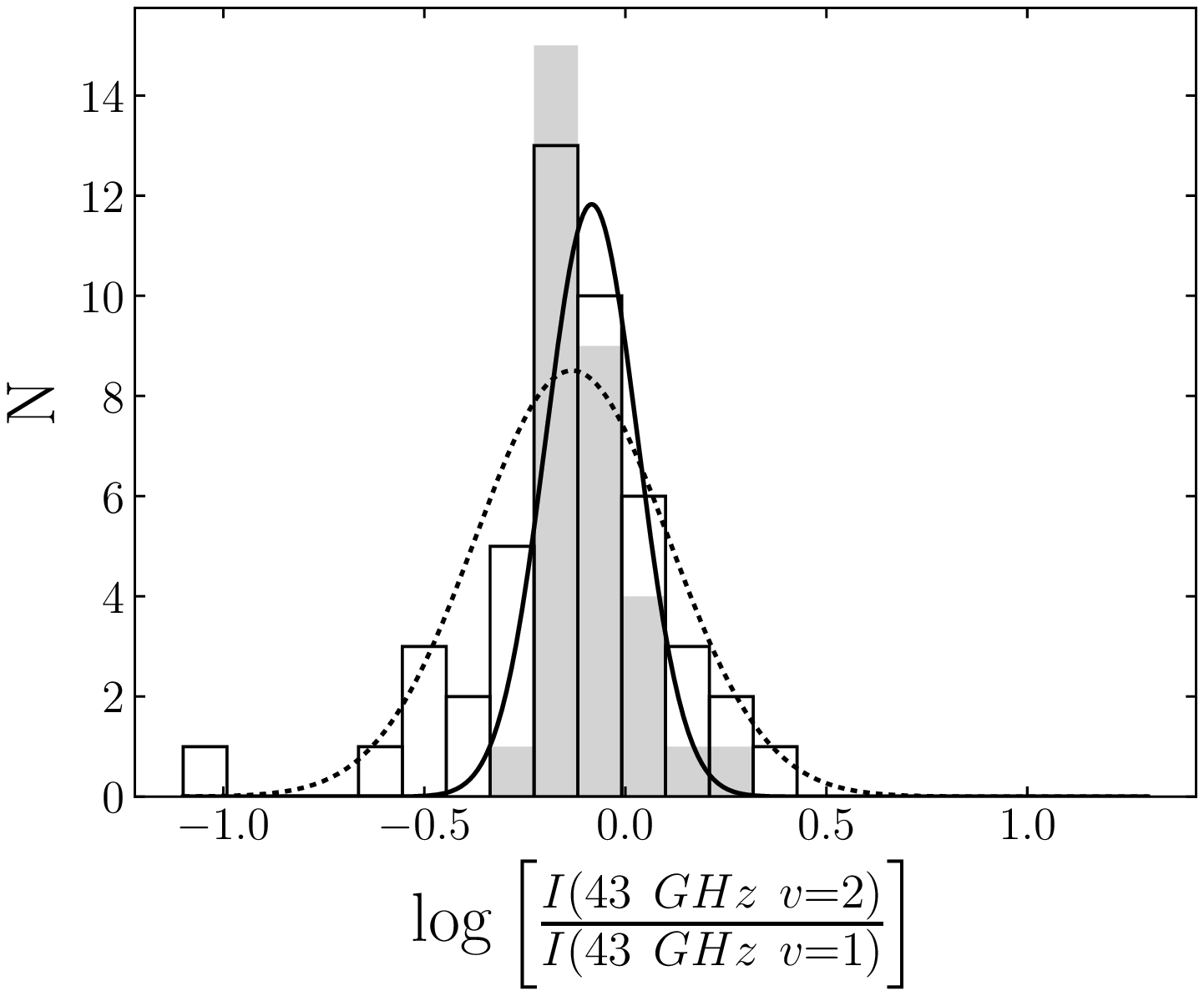}
\caption{Histogram of $\log\Big[\frac{ I(\textnormal{43~GHz}~v=2) }{ I(\textnormal{43~GHz}~v=1) }\Big]$ distributions for the populations where the 43 GHz $v$=3 line was detected in gray, and where the line was not detected are outlined. For sources with 43 GHz $v$=3 detected lines, the center and standard deviations of the distribution (solid line) are $-0.08\pm0.02$ and $0.11\pm0.01$, respectively. For sources without 43 GHz $v$=3 detected emission, the center and standard deviations of the distribution (dotted line) are $-0.13\pm0.04$ and $0.24\pm0.02$, respectively. Thus on average, the sources with 43 GHz $v$=3 detections, the emission from 43 $v$=1 \& 2 transitions are more tightly coupled than for sources without $v$=3 line emission.}
\label{fig_43_v2_v1_histogram_with_v3}
\end{figure}

Figure \ref{fig_43_v2_v1_scatter_with_v3} compares the 43 GHz $v$=2 to $v$=1 integrated flux densities and hints at a smaller dispersion for the subset of sources with $v$=3 line detections, which was suggested by the observations of \citet{1996AJ....111.1987C}. We quantify this smaller dispersion in Fig.\ \ref{fig_43_v2_v1_histogram_with_v3} by performing a maximum-likelihood regression of the $\log\Big[\frac{I(\textnormal{43~GHz}~v=2)}{I(\textnormal{43~GHz}~v=1)}\Big]$ distribution to a Gaussian distribution using \textit{fitdistr} resulting in a center at $-0.08\pm0.02$ ($-0.13\pm0.04$) with a standard deviation of $0.11\pm0.01$ ($0.24\pm0.02$) for the subset of sources with (without) 43 GHz $v$=3 line detections. The centers of the distributions for both subsets are consistent with the center for the full population (i.e. Fig.\ \ref{fig_consolidated_histogram}, center), but the dispersion of the 43 GHz $v$=3 detected subset is approximately half of that of the sources where the $v$=3 line is not detected. This supports the tighter relation between 43 GHz $v$=1 and 2 emission for sources with $v$=3 emission, first pointed out by \citet{1996AJ....111.1987C}. 

We also note that at the 99\% confidence level, the Anderson-Darling test cannot reject the null hypotheses that these distributions are Gaussian since it gives p-values of 0.015 and 0.03 for the distributions with and without $v$=3 line detections, respectively. Although we cannot statistically rule out a Gaussian distribution, the low p-values suggest that further study of this ratio may greatly aid our understanding of the pumping mechanisms and physical environments that lead to these emission processes.

Figure \ref{fig_43_v2_v1_scatter_with_v3} also demonstrates that none of the sources with detected 43 GHz $v$=3 lines are associated with weak 43 GHz $v$=1 or $v$=2 maser emission. Assuming that there is no selection effect toward nearby sources, they could be intrinsically brighter and be explained by the 43 GHz $v$=3 lines being associated with the brightest IR phase of the Mira stellar cycle. Therefore, the 43 GHz $v$=3 detected sample represents a class of stars with brighter peak 43 GHz $v$=1 maser emission relative to the 86 GHz $v$=1 emission. There is not enough evidence for the Wilcoxon signed-rank test to reject the null hypothesis that the relative line strengths are equal for the population where the 43 GHz $v$=3 line is not detected.

To conclude, the following trends with respect to the presence of the $v$=3 lines can be discerned: \textit{a}) most of the sources displaying $v$=3 detections have 43 GHz lines that are significantly brighter than 86 GHz $v$=1 lines, \textit{b}) there is a smaller difference between the 43 GHz $v$=1 and $v$=2 line brightnesses for sources with the $v$=3 detected than for those in which it was not detected, and \textit{c}) the 43 GHz $v$=3 detections tend to be associated with brighter 43 GHz $v$=1 and $v$=2 lines.

These results can be discussed within the framework of the models of \citet{2014A&A...565A.127D} and \citet{2002A&A...386..256H}, as they outline changes in the maser line intensities as a function of density. First, the results by \citet{2002A&A...386..256H} build on hydrodynamic pulsation models of Miras \citep{1988ApJ...329..299B, 1989_nato}, and show that the density in the region where the SiO masers occur (2 AU in their model) can change by two orders of magnitude or more from cycle to cycle ($n_{H_2}\approx 3\times 10^{9}-1\times 10^{12}$ cm$^{-3}$).  Second, their model demonstrates that the density in the maser regime can change by an order of magnitude over one individual cycle. This is important, as the results from \citet{1996AJ....111.1987C} indicate that the $v$=3 emission lines may be exclusively associated with the IR maximum stellar phase for Miras, and are therefore likely associated with a higher density due to the stellar pulsations. We lack IR and optical variability information for these sources which would allow us to confirm the proposed trend by \citet{1996AJ....111.1987C}. Third, \citet{2014A&A...565A.127D} show that there is a significant change in relative line ratios when the density changes, with diagnostic plots including the photon intensity of both the 43 and 86 GHz $v$=1, 2 and 3 line sets covering the above density regime. In particular, they argue that for sources where the $v$=3 line is observed, the density must be above a certain critical value.

\begin{figure}[t]
\figurenum{7}
\includegraphics[scale=0.13]{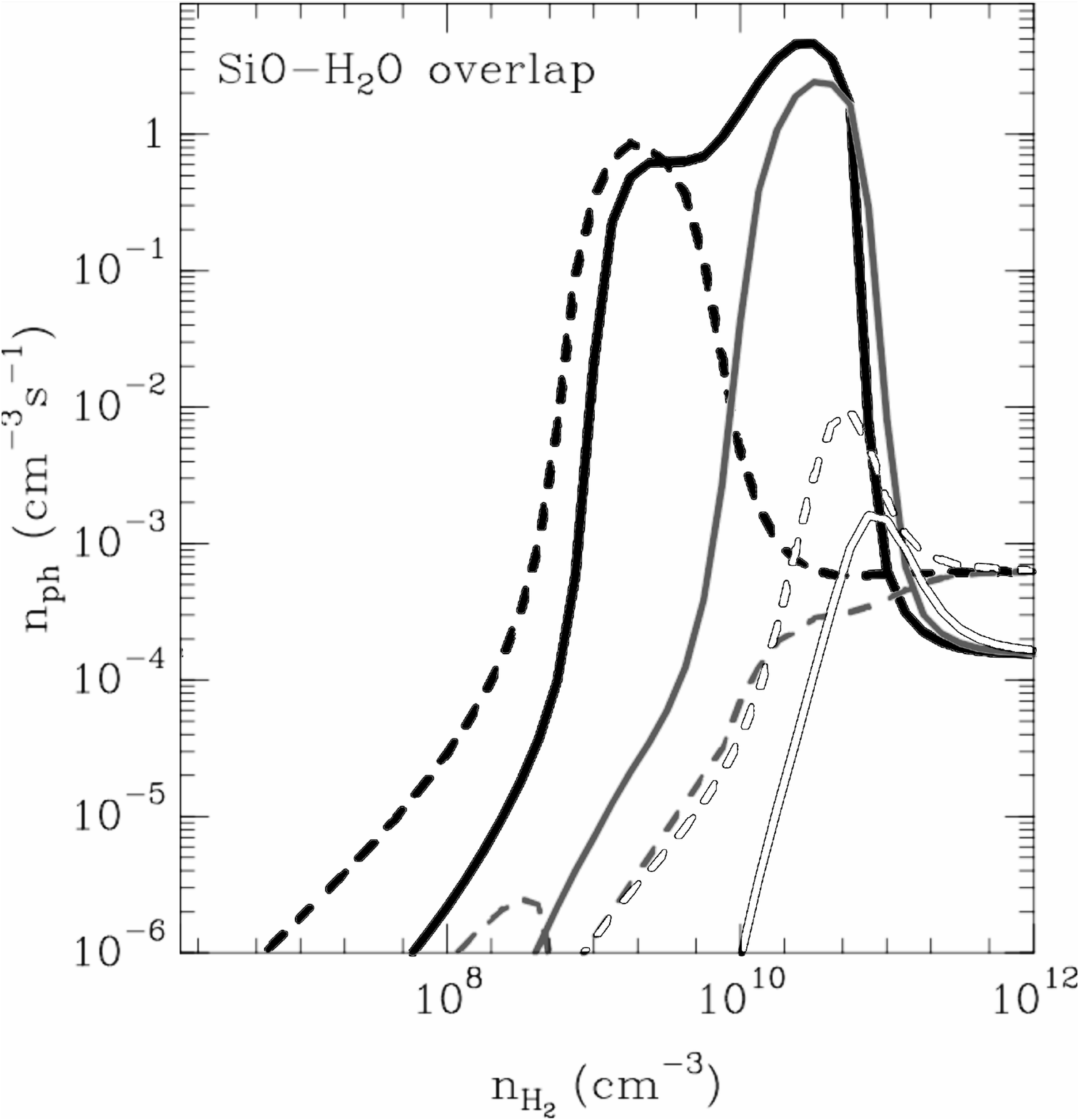}
\caption{Emitted photon density rate as a function of H$_2$ number density for SiO maser transitions, adapted from \citet{2014A&A...565A.127D}. Solid and dashed lines represent the 43 GHz and 86 GHz transitions, respectively, for radiative pumping models that include the SiO--H$_2$O line overlap. Black, gray and white represent the $v$=1, 2, and 3 states, respectively.}
\label{fig_desmurs_right_panel}
\end{figure}

Due to the significant lack of emission observed in the 86 GHz v=2 line, and to the success of the model for the pumping line overlap between SiO and H$_2$O first outlined by \citet{1981ApJ...247L..81O}, we will henceforth refer to Fig.\ \ref{fig_desmurs_right_panel}, which specifically includes this overlap effect. For the purpose of using their plot to discuss the observed $v$=3 results, we loosely define a high-density regime as $n_{H_2}>10^{10}$ cm$^{-3}$, and a low-density regime below this value. Comparing the model with the observed main $v$=3 trends \textit{a}-\textit{c} outlined above, we draw the following conclusions:

\renewcommand{\labelenumi}{\alph{enumi})}
\begin{enumerate}
     \item The $v$=3 lines are formed in the high-density regime. In this regime, near $n_{H_2} \sim 10^{10}$ cm$^{-3}$, the 43 GHz $v$=1 line is modeled to be brighter than the 86 GHz $v$=1 line. In the low-density regime, these two lines are on average more equal in strength. This is in full agreement with observational trend \textit{a}.
     \item In the high-density regime, the 43 GHz $v$=2 and $v$=1 line brightnesses are expected to follow each other closely and to be of near equal strength. In the low-density regime, there is a much larger distribution of the difference between the lines, with $v$=1 being the brightest, in agreement with the observed trend \textit{b}.
     \item The low-density regime supports the formation of $v$=1 and $v$=2 masers over a large range of brightnesses, including a range of weak masers at densities $< 10^9$cm$^{-3}$,  while the high-density regime does not support the range of weakest masers. This is in overall agreement with observed trend \textit{c}.  However, the observational trend \textit{c} may partly be due to sensitivity, as the $v$=3 line in general is expected to be weaker than the $v$=1 and 2, so if observed to the same sensitivity levels we would expect to detect the $v$=3 lines in either more nearby, or brighter sources
\end{enumerate}

In summary, the statistical differences between line ratios for detections versus non-detections in the 43 GHz $v$=3 line can serve as a useful probe of the density conditions in the circumstellar shell. The large spread in the $\log\Big[\frac{ I(\textnormal{43~GHz}~v=1) }{ I(\textnormal{86~GHz}~v=1) }\Big]$ and $\log\Big[ \frac{ I(\textnormal{43~GHz}~v=3) }{ I(\textnormal{43~GHz}~v=1) }\Big]$ distributions are easily explained by this model. Moreover, the model predicts that the 86 GHz $v$=3  line should be brighter than the 43 GHz $v$=3 line, which can be tested observationally.



\begin{figure*}[t]
\figurenum{8}
\includegraphics[scale=0.475]{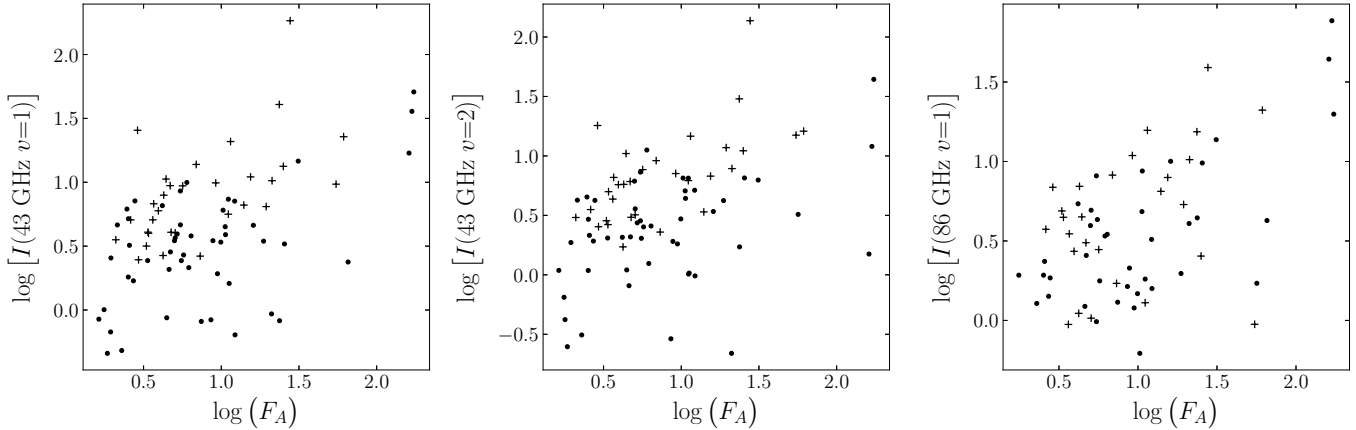}
\caption{Comparison between $I(\textnormal{43~GHz~}v=1)$, $I(\textnormal{43~GHz~}v=2)$, and $I(\textnormal{86~GHz~}v=1)$ versus $F_A$ (left, center and right, respectively). Crosses represent sources where 43 GHz $v$=3 emission was detected.}
\label{fig_consolidated_correlate_msx_a}
\end{figure*}

\subsection{Infrared Correlations}

The exact nature and the importance of infrared radiation providing at least part of the pumping energy of the SiO masers have long been discussed \citep{1974ApJ...194L..97K, 1976PASJ...28..307D, 1979ApJ...231..124C, 1981A&A...102...65B, 1984ApJ...284..751L, 1987A&A...175..164B, 1994A&A...285..953B, 2007ApJ...669..446N}. Both 4$\mu$m and 8$\mu$m photons are considered, as the 8$\mu$m could excite via the $\Delta v$=1 and the 4$\mu$m the $\Delta v$=2 transitions. Indeed, \citet{1987A&A...175..164B} demonstrated a correlation between the SiO maser intensity and 8$\mu$m intensity, and \citet{1996AJ....111.1987C} found a correlation between the maser intensity and 4$\mu$m intensity.

For the ATCA data, weak correlations\footnote{All correlations were tested using the Spearman's rank correlation test and only correlations at the 99\% significance level are considered.} are found between the integrated flux densities $I(\textnormal{43~GHz}~v=1)$, $I(\textnormal{43~GHz}~v=2)$ and $I(\textnormal{86~GHz}~v=1)$ and the MSX flux densities (not corrected for extinction or reddening) $F_A$, $F_C$, $F_D$ and $F_E$ measured in Jy. The strongest correlations are found with $F_A$ (8.28$\mu$m), giving correlation coefficients of 0.38, 0.40 and 0.39 for the 43 GHz $v=1$, $v=2$ and $v=3$ lines, respectively (see Fig.\ \ref{fig_consolidated_correlate_msx_a}). The strength of the correlations decreases toward longer MSX wavelengths, to the point where the correlation between 86 GHz $v=1$ against $F_E$ falls below the 99\% confidence level. That these correlations are stronger with $F_A$ is expected, given that the wavelength of Band A matches the wavelengths of the fundamental vibrational-rotational pumping transition, and that the photon energies of the longer wavelength bands are insufficient to contribute to the pumping \citep[e.g.,][and references above]{1996A&A...314..883B}.

The 43 GHz correlations are notably strengthened if only sources with detected 43 GHz $v=3$ emission are considered, resulting in correlation coefficients of 0.60 and 0.62 for the $v=1$ and $v=2$ lines respectively. No statistically significant correlation is found between the 86 GHz $v=1$ line and the MSX fluxes using the 43 GHz $v=3$ detected sample, which may be partially due to the smaller sample size compared to that of the 43 GHz v=1 and v=2 lines.

\begin{figure}
\figurenum{9}
\includegraphics[scale=0.5]{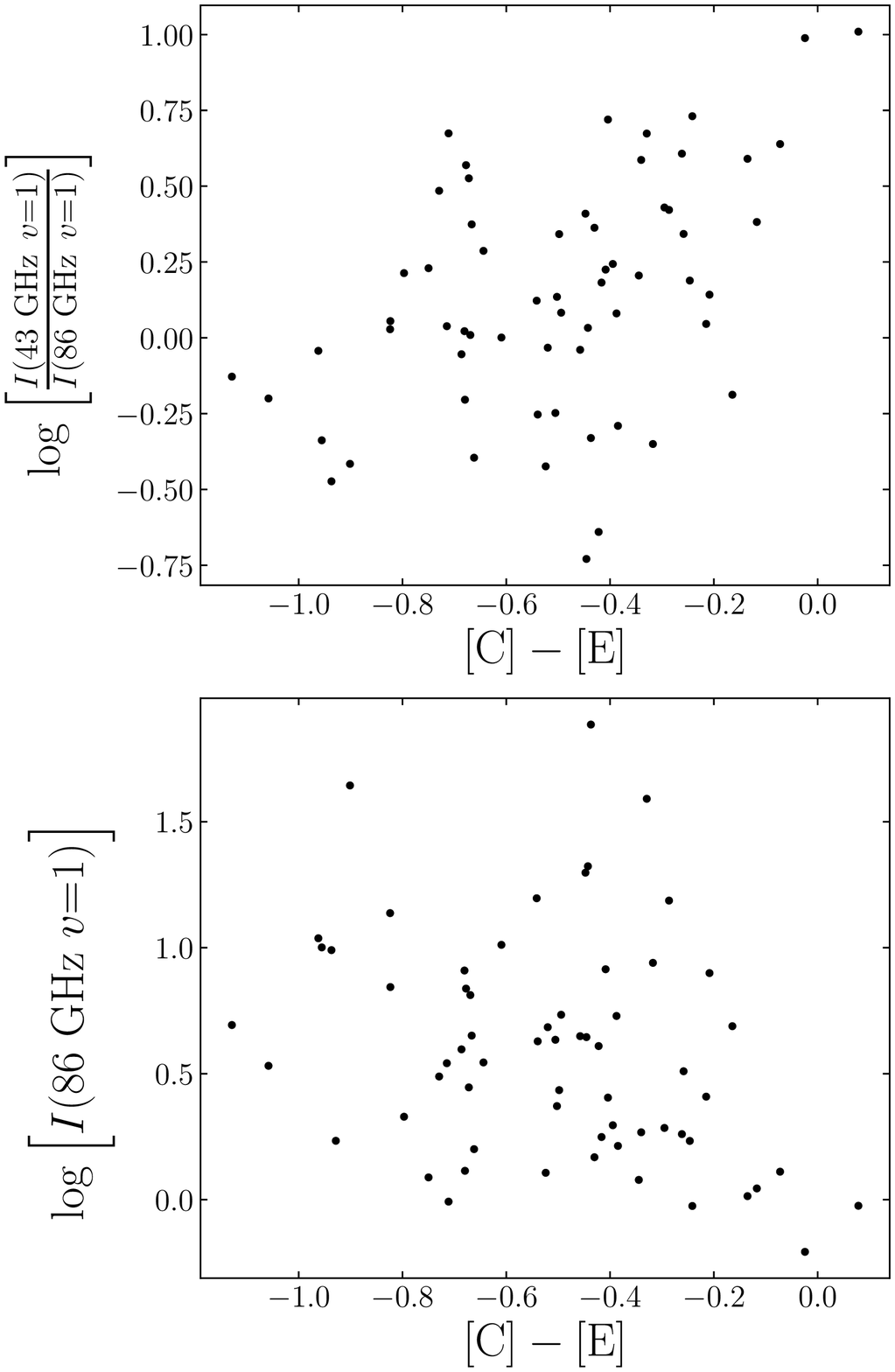}
\caption{Top: Comparison between $\frac{I(\textnormal{43~GHz~}v=1) }{ I(\textnormal{86~GHz~}v=1) }$ and [C]--[E]. Bottom: Comparison between $I(\textnormal{86~GHz~}v=1)$ and [C]--[E].}
\label{fig_consolidated_correlate_msx_e_c}
\end{figure}

No correlations are found between the integrated flux densities or line ratios and the MSX [A]--[D] color, which is the primary color used to select BAaDE sources \citep{in...prep}. A weak correlation with correlation coefficient of 0.43 (0.38) is found between $\frac{I(\textnormal{43~GHz}~v=1)}{I(\textnormal{86~GHz}~v=1)}$ and [C]--[E] ([A]--[E]), see Fig.\ \ref{fig_consolidated_correlate_msx_e_c}, which are the alternative colors BAaDE sources were chosen from. However, this correlation is markedly diminished if only a few points are removed from the plot. This correlation may be mostly due to an anti-correlation between $I(\textnormal{86~GHz}~v=1)$ and [C]--[E] (correlation coefficient of $-0.33$, see bottom of Fig.\ \ref{fig_consolidated_correlate_msx_e_c}) but is similarly diminished with the removal of a few data points.
Unlike \citet{2007ApJ...669..446N}, no statistically significant correlations are found between $\frac{I(\textnormal{43~GHz}~v=2)}{I(\textnormal{43~GHz}~v=1)}$ and [C]--[E]. Their data include redder sources (e.g. $-1<[\textnormal{C}]-[\textnormal{E}]<1$ whereas the BAaDE ATCA sample is restricted to $-1< [\textnormal{C}]-[\textnormal{E}]< 0$) so a stronger relation could reside in the redder region, or by sampling a larger range, the correlation could be more prominent.

Finding statistically significant correlations with infrared colors in the MSX band may require larger samples, as the flux densities in the MSX bands vary with stellar phase, and are not observed simultaneously with the SiO maser data. In addition, the presence of the silicate dust feature at 9.7$\mu$m could affect the $F_A$ and possibly the $F_C$ values. Moreover, the BAaDE ATCA sample selection is biased toward the brighter SiO maser sources, thereby limiting the dynamic range when searching for a correlation between the maser flux density and other variables. A more robust statistical study will be performed with observations from the BAaDE survey which will contain thousands of simultaneous observations of the 43 GHz $v$=0, 1, 2 \& 3 using the VLA and thousands of simultaneous observations of the 86 GHz $v$=0, 1 \& 2 lines.

\section{Summary}
The BAaDE project uses both 43 and 86 GHz SiO maser transitions to
obtain line-of-sight velocities of IR-selected late-type stars
throughout the Galactic plane. Our ATCA observations indicate that the
relative strength of the 43 GHz $v$=1 line is on average 1.3 times
stronger than the 86 GHz $v$=1 line. This means that by using the 86
GHz transition with ALMA for the far side of the bulge and bar to
derive a velocity using the same observing rms as with the VLA observing the 43 GHz transition elsewhere in the bulge and bar, a fraction of the weaker SiO maser population will
be missed. However, due to the order-of-magnitude spread of line ratios, we expect the relative shortfall of detections on the far side of the Galactic plane to be less than 30\%. 
VLA observing the 43 GHz transition elsewhere in the bulge and bar, a fraction of the weaker SiO maser population will be missed. In order to achieve the same overall source detection rate at both 43 and 86 GHz, the 86 GHz integration time would have to be increased by a factor of 1.7.

There is evidence that the 43 GHz $v$=2 line on average $\sim23\%$ weaker than the $v$=1 line, with a much tighter correlation than
what is observed between the 43 and 86 GHz $v$=1 lines and the 43 GHz
$v$=1 and $v$=3 lines.  Furthermore, the 43 GHz $v$=3 line is typically only about 10\% as strong as the 43 GHz $v$=1 line.

The presence of a detectable 43 GHz $v$=3 line alters the statistics \citep[as was previously reported by][]{1996AJ....111.1987C}, consistent with the SiO masers displaying $v$=3
emission arising in a denser regime in the circumstellar shell compared to those without. The
density difference may be due to the distance from the star, and/or to density
variations caused by the stellar pulsations. As a result, there is no statistical difference between the relative line strengths of the 43 and 86 GHz $v$=1 lines for sources no detectable $v$=3 emission; however, there is strong evidence that sources with the $v$=3 emission will on average have 43 GHz $v$=1 emission that is nearly twice as luminous as the 86 GHz $v$=1 emission. 

The analysis is based on the results from a set of the brightest SiO
masers detected in the BAaDE survey, and it is therefore possible that statistical correlations will strengthen, or possibly weaken, when a
larger range of SiO maser flux densities is included. Thus, although the full
BAaDE sample does not contain simultaneous 43 GHz and 86 GHz
observations, it will provide thousands of ratios of simultaneously observed lines from different vibrational states within each of the 43 and 86 GHz bands. 

By choosing from among the brightest sources, these may also be the closest and may consist of a limited range of ages and metallicities. We cannot rule out that our correlations and results may change for populations at different regions in the Galactic Plane.

An examination of the variability of these sources combining the initial VLA and ALMA detections with the ATCA results is planned for a future paper, as well as a study on the stellar phases and periods and their respective association with the SiO maser transitions. To further explore the pumping models, it would also be interesting to include the 86 GHz $v$=3 line in our observations.

\acknowledgments
The authors would like to thank Jean-Francois Desmurs for providing Fig.\ \ref{fig_desmurs_right_panel}.

The Australia Telescope Compact Array is part of the Australia Telescope National Facility which is funded by the Australian Government for operation as a National Facility managed by CSIRO.

The National Radio Astronomy Observatory is a facility of the National Science Foundation operated under cooperative agreement by Associated Universities, Inc.

This material is based upon work supported by the National Science Foundation under Grant Number 1517970 to UNM and 1518271 to UCLA. Any opinions, findings, and conclusions or recommendations expressed in this material are those of the authors and do not necessarily reflect the views of the National Science Foundation.

\facility{ATCA}
\software{MASS \citep{ISBN0-387-95457-0}, MIRIAD \citep{1995ASPC...77..433S}}

\clearpage

\begin{longrotatetable}
\begin{deluxetable*}{llllrrrrrrrrr}
\tablecaption{Integrated flux densities and brightness ratios of BAaDE SiO Masers \label{maser_summary_table}}
\tabletypesize{\scriptsize}
\tablehead{
\colhead{BAaDE} & 
\colhead{Alternative} & 
\colhead{Survey} & 
\colhead{Observing} & 
\colhead{$I(\textnormal{43~GHz}~v=1)$} &
\colhead{$I(\textnormal{43~GHz}~v=2)$} &
\colhead{$I(\textnormal{43~GHz}~v=3)$} &
\colhead{$I(\textnormal{86~GHz}~v=0)$} &
\colhead{$I(\textnormal{86~GHz}~v=1)$} &
\colhead{$I(\textnormal{86~GHz}~v=2)$} &
\colhead{} & 
\colhead{} &
\colhead{}
\\
\colhead{Name} & 
\colhead{Name} & 
\colhead{Sample} & 
\colhead{Date} & 
\colhead{(Jy~km~s$^{-1}$)} &
\colhead{(Jy~km~s$^{-1}$)} &
\colhead{(Jy~km~s$^{-1}$)} &
\colhead{(Jy~km~s$^{-1}$)} &
\colhead{(Jy~km~s$^{-1}$)} &
\colhead{(Jy~km~s$^{-1}$)} &
\colhead{$\frac{I(\textnormal{43~GHz}~v=1)}{I(\textnormal{43~GHz}~v=1) }$} & 
\colhead{$\frac{I(\textnormal{43~GHz}~v=2)}{I(\textnormal{43~GHz}~v=1) }$} &
\colhead{$\frac{I(\textnormal{43~GHz}~v=3)}{I(\textnormal{43~GHz}~v=1) }$}
}

\startdata
ad3a-06173 &  & VLA & 20160904 &        3.55 &        3.04 &        0.59 &  <      0.55 & <      0.77$^{c}$ & <      0.86 &     \nodata &        0.86 &        0.17 \\ 
ad3a-06187 &  & VLA & 20160728 &        2.70 &        2.53 & <      0.20 & <      0.51 &        1.77 & <      0.53 &        1.52 &        0.94 &     \nodata \\ 
ad3a-06197 &  & VLA & 20160903 &        0.67 & <      0.30$^{a}$ & <      0.29 & <      1.09 & <      1.26$^{c}$ & <      1.17 &     \nodata &     \nodata &     \nodata \\ 
ad3a-06199 &  & VLA & 20160903 &        4.06 &        3.05 &        0.35 & <      0.98 & <      1.13$^{c}$ & <      1.02 &     \nodata &        0.75 &        0.09 \\ 
ad3a-06214 &  & VLA & 20160904 & <      0.27$^{a}$ & <      0.25$^{a}$ & <      0.27 & <      0.49 & <      0.65$^{c}$ & <      0.72 &     \nodata &     \nodata &     \nodata \\ 
ad3a-06218 & V5357 Sgr & VLA & 20160727 &       14.64 &        6.26 & <      0.36 & <      0.32 &       13.72 & <      0.54 &        1.07 &        0.43 &     \nodata \\ 
ad3a-06239 &  & VLA & 20160903 &        1.81 &        1.09 & <      0.36 & <      1.85 & <      2.03$^{c}$ & <      1.90 &     \nodata &        0.60 &     \nodata \\ 
ad3a-06253 &  & VLA & 20160903 &        7.37 &        6.50 & <      0.27 & <      1.05 &        1.82 & <      1.13 &        4.05 &        0.88 &     \nodata \\ 
ad3a-06272 &  & VLA & 20160904 &        7.93 &        5.77 &        0.67 & <      0.55 &        6.98 & <      0.84 &        1.14 &        0.73 &        0.08 \\ 
ad3a-06275 &  & VLA & 20160729 & <      0.25$^{a}$ &        1.01 & <      0.23 & <      0.54 & <      0.60$^{c}$ & <      0.58 &     \nodata &     \nodata &     \nodata \\ 
ad3a-06287 &  & VLA & 20160729 &        6.17 &        4.51 & <      0.31 & <      0.52 & <      0.58$^{c}$ & <      0.52 &     \nodata &        0.73 &     \nodata \\ 
ad3a-06300 &  & VLA & 20160727 &        9.37 &        7.67 &        1.11 & <      0.28 &        2.79 & <      0.46 &        3.36 &        0.82 &        0.12 \\ 
ad3a-06301 &  & VLA & 20160908 & <      0.31$^{a}$ &        0.42 & <      0.27 & <      0.99 & <      1.01$^{c}$ & <      1.02 &     \nodata &     \nodata &     \nodata \\ 
ad3a-06315 &  & VLA & 20160729 &        0.64 &        0.98 & <      0.29 & <      0.76 &        1.59 & <      0.73 &        0.40 &        1.53 &     \nodata \\ 
ad3a-06338 &  & VLA & 20160908 & <      0.31$^{a}$ & <      0.31$^{a}$ & <      0.31 & <      0.96 & <      1.06$^{c}$ & <      0.98 &     \nodata &     \nodata &     \nodata \\ 
ad3a-06342 &  & VLA & 20160729 &        2.67 &        1.72 &        0.51 & <      0.63 &        1.11 & <      0.66 &        2.41 &        0.64 &        0.19 \\ 
ad3a-06353 &  & VLA & 20160728 &        4.61 &        3.41 & <      0.21 & <      0.53 &       10.03 & <      0.56 &        0.46 &        0.74 &     \nodata \\ 
ad3a-06360 &  & VLA & 20160903 &        2.55 &        1.87 & <      0.31 & <      1.11 & <      1.31$^{c}$ & <      1.16 &     \nodata &        0.73 &     \nodata \\ 
ad3a-06364 &  & VLA & 20160904 &        4.07 &        2.64 &        0.56 & <      0.54 &        4.46 & <      0.75 &        0.91 &        0.65 &        0.14 \\ 
ad3a-06375 &  & VLA & 20160729 &        2.64 &        2.29 &        0.73 & <      0.70 &        1.71 & <      0.72 &        1.54 &        0.87 &        0.28 \\ 
ad3a-06378 &  & VLA & 20160728 &        0.87 &        1.10 & <      0.30 & <      0.55 & <      0.65$^{c}$ & <      0.62 &     \nodata &        1.27 &     \nodata \\ 
ad3a-06391 &  & VLA & 20160904 &        3.94 &        2.74 & <      0.27 & <      0.56 & <      0.71$^{c}$ & <      0.84 &     \nodata &        0.69 &     \nodata \\ 
ad3a-06413 &  & VLA & 20160728 &        3.21 &        2.15 & <      0.20 & <      0.54 &        2.35 & <      0.58 &        1.36 &        0.67 &     \nodata \\ 
ad3a-06417 &  & VLA & 20160728 &       10.60 &       10.47 &        1.42 & <      0.48 &        4.48 & <      0.51 &        2.37 &        0.99 &        0.13 \\ 
ad3a-06420 & V5410 Sgr & VLA & 20160904 &        3.99 &        5.00 &        0.63 & <      0.54 & <      0.70$^{c}$ & <      0.84 &     \nodata &        1.25 &        0.16 \\ 
ad3a-06429 &  & VLA & 20160729 &        2.44 &        2.04 & <      0.25 & <      0.66 & <      0.70$^{c}$ & <      0.68 &     \nodata &        0.84 &     \nodata \\ 
ad3a-06431 & V4699 Sgr & VLA & 20160727 &        3.80 &        2.57 & <      0.33 & <      0.34 &        3.48 & <      0.60 &        1.09 &        0.68 &     \nodata \\ 
ad3a-06433 &  & VLA & 20160728 &        5.18 &        2.93 & <      0.22 & <      0.48 &        1.93 & <      0.48 &        2.69 &        0.57 &     \nodata \\ 
ad3a-06438 &  & VLA & 20160727 &       13.33 &       11.04 &        0.63 & <      0.11 &        2.54 & <      0.18 &        5.24 &        0.83 &        0.05 \\ 
ad3a-06484 & V1946 Sgr & VLA & 20160728 &        9.41 &        6.09 &        0.32 & <      0.59 &        3.08 & <      0.61 &        3.05 &        0.65 &        0.03 \\ 
ad3a-06576 &  & VLA & 20160729 &        0.85 &        1.09 & <      0.21 & <      0.48 & <      0.54$^{c}$ & <      0.49 &     \nodata &        1.29 &     \nodata \\ 
ad3a-06653 &  & VLA & 20160729 &        4.02 &        3.19 &        0.82 & <      0.53 &        1.03 & <      0.56 &        3.89 &        0.79 &        0.20 \\ 
ad3a-06662 &  & VLA & 20160903 &        4.63 &        4.25 & <      0.37 & <      1.41 & <      1.46$^{c}$ & <      1.39 &     \nodata &        0.92 &     \nodata \\ 
ad3a-06706 &  & VLA & 20160728 &        5.62 &        6.20 &        1.05 & <      0.50 &        1.29 & <      0.52 &        4.35 &        1.10 &        0.19 \\ 
ad3a-25844 &  & ALMA & 20160904 &        0.46 &        0.25 & <      0.26 & <      0.46 & <      0.69$^{d}$ & <      0.76 &     \nodata &        0.54 &     \nodata \\ 
ad3a-25852 &  & ALMA & 20160727 &       20.84 &       14.66 &        0.80 & <      0.45 &       15.71 & <      0.80 &        1.33 &        0.70 &        0.04 \\ 
ad3a-25928 &  & ALMA & 20160725 &        6.45 &       11.74 &        1.73 & <      0.30 &        5.36 & <      0.47 &        1.20 &        1.82 &        0.27 \\ 
ad3a-25950 &  & ALMA & 20160908 &        0.84 &        0.29 & <      0.24 & <      0.90 &        1.63 & <      0.93 &        0.51 &        0.35 &     \nodata \\ 
ad3a-25969 &  & ALMA & 20160908 &        0.93 &        0.22 & <      0.23 & <      0.83 &        4.07 &        1.76 &        0.23 &        0.23 &     \nodata \\ 
ad3a-26000 &  & ALMA & 20160904 &        1.01 &        0.65 & <      0.27 & <      0.46 &        1.93 & <      0.75 &        0.52 &        0.64 &     \nodata \\ 
ad3a-26038 &  & ALMA & 20160725 &        2.38 & <      0.22$^{b}$ & <      0.24 & <      0.33 &        4.25 & <      0.48 &        0.56 &     \nodata &     \nodata \\ 
ad3a-26052 &  & ALMA & 20160727 &        3.29 &        6.53 & <      0.23 & <      0.55 &        9.78 &       15.10 &        0.34 &        1.98 &     \nodata \\ 
ad3a-26104 &  & ALMA & 20160727 &        8.54 &        7.31 & <      0.33 & <      0.44 &        8.12 & <      0.72 &        1.05 &        0.86 &     \nodata \\ 
ad3a-26116 &  & ALMA & 20160728 &      183.96 &      137.06 &       12.23 & <      0.72 &       39.01 & <      0.78 &        4.72 &        0.75 &        0.07 \\ 
ad3a-26121 &  & ALMA & 20160725 &       25.54 &       18.02 &        0.90 & <      0.27 &        6.88 & <      0.48 &        3.71 &        0.71 &        0.04 \\ 
ad3a-26136 &  & ALMA & 20160728 &        6.78 &        6.59 &        0.76 & <      0.38 &        3.51 & <      0.37 &        1.94 &        0.97 &        0.11 \\ 
ad3a-26156 &  & ALMA & 20160728 &       50.97 &       44.05 & <      0.44 & <      1.39 &       19.86 & <      0.93 &        2.57 &        0.86 &     \nodata \\ 
ad3a-26234 &  & ALMA & 20160908 &        0.48 &        0.31 & <      0.23 & <      0.92 &        1.28 & <      0.91 &        0.38 &        0.65 &     \nodata \\ 
ad3a-26256 &  & ALMA & 20160727 &        4.49 &        5.08 & <      0.23 & <      0.46 &        4.84 & <      0.82 &        0.93 &        1.13 &     \nodata \\ 
ad3a-26294 &  & ALMA & 20160728 &        4.64 &        2.85 & <      0.25 & <      0.31 &        0.98 & <      0.34 &        4.72 &        0.61 &     \nodata \\ 
ad3a-26295 &  & ALMA & 20160908 &        3.46 &        4.21 & <      0.27 & <      1.00 &        1.97 & <      0.96 &        1.75 &        1.22 &     \nodata \\ 
ad3a-26300 &  & ALMA & 20160904 &        5.98 &        5.73 &        0.43 & <      0.60 &        2.72 & <      0.94 &        2.20 &        0.96 &        0.07 \\ 
ad3a-26315 &  & ALMA & 20160729 &        1.92 &        1.82 & <      0.22 & <      0.29 &        1.20 & <      0.28 &        1.60 &        0.95 &     \nodata \\ 
ad3a-26322 &  & ALMA & 20160727 &        3.89 &        4.39 & <      0.25 & <      0.46 &        8.71 & <      0.82 &        0.45 &        1.13 &     \nodata \\ 
ad3a-26339 &  & ALMA & 20160729 &        0.81 & <      0.21$^{b}$ & <      0.21 & <      0.21 &        1.30 & <      0.21 &        0.62 &     \nodata &     \nodata \\ 
ad3a-26342 &  & ALMA & 20160725 &       35.92 &       12.01 & <      0.23 & <      0.94 &       76.87 & <      1.22 &        0.47 &        0.33 &     \nodata \\ 
ad3a-26359 & V0388 Sco & ALMA & 20160904 &        3.49 &        1.91 & <      0.27 & <      0.51 &        2.13 & <      0.82 &        1.63 &        0.55 &     \nodata \\ 
ad3a-26431 &  & ALMA & 20160908 &        1.61 &        1.03 & <      0.25 & <      0.88 & <      1.05$^{d}$ & <      0.91 &     \nodata &        0.64 &     \nodata \\ 
ad3a-26448 &  & ALMA & 20160725 &       40.63 &       30.16 &        3.51 & <      0.49 &       15.39 & <      0.73 &        2.64 &        0.74 &        0.09 \\ 
ad3a-26466 &  & ALMA & 20160729 &        3.40 &        2.95 & <      0.22 & <      0.23 &        1.47 & <      0.24 &        2.31 &        0.87 &     \nodata \\ 
ad3a-26488 &  & ALMA & 20160727 &       11.01 &        6.77 &        0.58 & <      0.47 &        7.93 & <      0.77 &        1.39 &        0.61 &        0.05 \\ 
ad3a-26538 &  & ALMA & 20160908 &        2.47 &        2.54 &        0.43 & <      0.89 & <      0.99$^{d}$ & <      0.90 &     \nodata &        1.03 &        0.17 \\ 
ad3a-26590 & RT Sco & ALMA & 20160725 &       16.92 &        1.50 & <      0.15 &        0.81 &       44.06 &        1.61 &        0.38 &        0.09 &     \nodata \\ 
ad3a-26662 &  & ALMA & 20160904 &        9.96 &       11.21 & <      0.28 & <      0.47 & <      0.71$^{d}$ & <      0.77 &     \nodata &        1.12 &     \nodata \\ 
ad3a-26682 &  & ALMA & 20160727 &        3.49 &        6.14 & <      0.30 & <      0.44 &        3.95 & <      0.76 &        0.88 &        1.76 &     \nodata \\ 
ad3a-26756 &  & ALMA & 20160729 &        2.85 &        2.09 & <      0.22 & <      0.29 &        2.56 & <      0.29 &        1.11 &        0.73 &     \nodata \\ 
ad3a-26789 &  & ALMA & 20160727 &        3.17 &        2.85 &        0.34 & <      0.43 &        4.88 & <      0.80 &        0.65 &        0.90 &        0.11 \\ 
ad3a-26792 &  & ALMA & 20160727 & <      0.23$^{b}$ &        3.22 & <      0.22 & <      0.53 &        1.71 & <      0.91 &     \nodata &     \nodata &     \nodata \\ 
ad3a-26796 &  & ALMA & 20160729 &        7.12 &        5.14 & <      0.21 & <      0.22 &        3.23 & <      0.25 &        2.20 &        0.72 &     \nodata \\ 
ad3a-26805 &  & ALMA & 20160904 &        1.69 &        1.92 & <      0.27 & <      0.57 &        1.42 & <      0.92 &        1.19 &        1.14 &     \nodata \\ 
ad3a-26811 &  & ALMA & 20160725 &        3.68 &        3.59 & <      0.18 & <      0.31 &        4.94 & <      0.52 &        0.74 &        0.98 &     \nodata \\ 
ad3a-26821 &  & ALMA & 20160727 &        2.44 &        2.03 & <      0.23 & <      0.48 &        4.31 & <      0.81 &        0.57 &        0.83 &     \nodata \\ 
ad3a-26835 &  & ALMA & 20160729 &        7.15 &        4.23 & <      0.21 & <      0.21 &        1.85 & <      0.21 &        3.86 &        0.59 &     \nodata \\ 
ad3a-26872 &  & ALMA & 20160729 &        6.04 &        6.51 & <      0.21 & <      0.22 &        0.62 & <      0.25 &        9.73 &        1.08 &     \nodata \\ 
ad3a-26936 &  & ALMA & 20160908 &        2.15 &        1.24 & <      0.22 & <      0.87 &        3.40 & <      0.94 &        0.63 &        0.58 &     \nodata \\ 
ad3a-26998 &  & ALMA & 20160908 &        6.63 &        3.38 &        1.02 & <      1.16 &        6.49 & <      1.15 &        1.02 &        0.51 &        0.15 \\ 
ad3a-27001 &  & ALMA & 20160904 &        5.07 &        3.54 &        1.34 & <      0.49 &        3.75 & <      0.80 &        1.35 &        0.70 &        0.26 \\ 
ad3a-27027 &  & ALMA & 20160728 &        2.08 &        0.81 & <      0.26 & <      0.36 &        1.23 & <      0.35 &        1.70 &        0.39 &     \nodata \\ 
ad3a-27030 &  & ALMA & 20160729 &        5.08 &        4.34 &        0.36 & <      0.22 &        0.94 & <      0.24 &        5.38 &        0.85 &        0.07 \\ 
ad3a-27041 &  & ALMA & 20160728 &       13.79 &        9.13 &        0.61 & <      0.42 &        8.21 & <      0.46 &        1.68 &        0.66 &        0.04 \\ 
ad3a-27099 &  & ALMA & 20160725 &       22.72 &       16.15 &        0.43 & <      0.41 &       21.06 &        3.23 &        1.08 &        0.71 &        0.02 \\ 
ad3a-27103 &  & ALMA & 20160727 &        0.82 &        1.72 & <      0.23 & <      0.42 &        4.42 & <      0.76 &        0.19 &        2.08 &     \nodata \\ 
ad3a-27107 &  & ALMA & 20160908 &        9.88 &        7.11 &        0.91 & <      0.97 &       10.90 & <      0.99 &        0.91 &        0.72 &        0.09 \\ 
ad3a-27122 &  & ALMA & 20160727 &       10.29 &        7.82 &        0.57 & <      0.50 &       10.27 & <      0.76 &        1.00 &        0.76 &        0.06 \\ 
ae3a-00554 &  & ALMA & 20160904 &        6.56 &        2.07 & <      0.27 & <      0.52 &        5.42 & <      0.78 &        1.21 &        0.32 &     \nodata \\ 
ce3a-00362 & V1013 Sco & ALMA & 20160908 &        9.67 &       14.94 &        0.49 & <      0.85 &        0.95 & <      0.92 &       10.22 &        1.55 &        0.05 \\ 
\enddata
\tablecomments{Alternative names are known variables within 2" of the 2MASS position for these sources. Survey sample refers to the observatory for which the source was originally detected as part of the BAaDE survey. Integrated values are calculated over 3~km~s$^{-1}$. Upper limits were calculated using $I(i) \le 5 \sigma n^{1/2} \Delta v$ where $\sigma$ is the rms noise, n is the number of bins integrated over (i.e., three) and $\Delta v$ is the velocity resolution of the spectra (i.e., one km~s$^{-1}$) (Nyman \& Olofsson 1986).}
\tablenotetext{a}{Could indicate source variability since the upper limit is lower than the previously detected flux density from our 43 GHz campaign.}
\tablenotetext{b}{No previous 43 GHz detection since this was observed as part of our 86 GHz campaign.}
\tablenotetext{c}{No previous 86 GHz detection since this was observed as part of our 43 GHz campaign.}
\tablenotetext{d}{The ATCA upper limit is consistent with previous 86 GHz observations.}
\tablenotetext{e}{Could indicate source variability since the upper limit is lower than the previously detected flux density from our 86 GHz campaign.}
\label{tab_results}
\end{deluxetable*}
\end{longrotatetable}

\appendix
\section{BAaDE ATCA Source List}
\startlongtable
\begin{deluxetable*}{lrrrr}
\tablecaption{Coordinates of BAaDE ATCA Sample \label{source_coordinates_table}}
\tabletypesize{\scriptsize}
\tablehead{
\colhead{BAaDE} & 
\colhead{2MASS R.A.} & 
\colhead{2MASS Dec.} & 
\colhead{MSX R.A.} & 
\colhead{MSX Dec.}
\\
\colhead{Name} & 
\colhead{(J2000)} & 
\colhead{(J2000)} & 
\colhead{(J2000)} & 
\colhead{(J2000)}
}

\startdata
ad3a-06173 & 17~54~10.38 & -24~23~35.9 & 17~54~10.39 & -24~23~34.4 \\
ad3a-06187 & 18~05~11.84 & -24~21~15.9 & 18~05~11.83 & -24~21~16.2 \\
ad3a-06197 & 17~37~28.59 & -24~20~21.7 & 17~37~28.61 & -24~20~20.8 \\
ad3a-06199 & 17~55~29.53 & -24~19~57.0 & 17~55~29.59 & -24~19~55.6 \\
ad3a-06214 & 17~40~27.26 & -24~16~52.7 & 17~40~27.41 & -24~16~50.2 \\
ad3a-06218 & 18~00~32.79 & -24~16~19.0 & 18~00~32.78 & -24~16~18.8 \\
ad3a-06239 & 18~01~25.55 & -24~13~00.1 & 18~01~25.58 & -24~12~59.0 \\
ad3a-06253 & 18~07~50.37 & -24~10~21.6 & 18~07~50.42 & -24~10~22.4 \\
ad3a-06272 & 17~59~16.64 & -24~07~04.9 & 17~59~16.70 & -24~07~04.4 \\
ad3a-06275 & 17~55~27.54 & -24~06~43.5 & 17~55~27.67 & -24~06~40.7 \\
ad3a-06287 & 18~05~42.13 & -24~05~08.3 & 18~05~42.17 & -24~05~08.2 \\
ad3a-06300 & 18~02~02.93 & -24~03~30.9 & 18~02~02.90 & -24~03~29.9 \\
ad3a-06301 & 18~15~54.16 & -24~03~27.3 & 18~15~54.26 & -24~03~26.3 \\
ad3a-06315 & 18~08~28.61 & -23~59~20.1 & 18~08~28.68 & -23~59~20.0 \\
ad3a-06338 & 17~41~53.27 & -23~54~52.2 & 17~41~53.52 & -23~54~49.3 \\
ad3a-06342 & 18~16~55.44 & -23~53~35.7 & 18~16~55.58 & -23~53~35.2 \\
ad3a-06353 & 18~08~22.75 & -23~51~26.6 & 18~08~22.78 & -23~51~25.9 \\
ad3a-06360 & 17~51~27.33 & -23~50~36.2 & 17~51~27.34 & -23~50~36.2 \\
ad3a-06364 & 18~06~26.26 & -23~49~42.1 & 18~06~26.28 & -23~49~41.9 \\
ad3a-06375 & 18~09~37.10 & -23~47~39.5 & 18~09~37.18 & -23~47~38.8 \\
ad3a-06378 & 18~00~26.85 & -23~47~26.4 & 18~00~26.90 & -23~47~25.4 \\
ad3a-06391 & 17~34~21.46 & -23~46~08.3 & 17~34~21.29 & -23~46~07.7 \\
ad3a-06413 & 17~53~47.42 & -23~43~13.9 & 17~53~47.47 & -23~43~13.4 \\
ad3a-06417 & 17~59~25.33 & -23~42~42.7 & 17~59~25.32 & -23~42~42.8 \\
ad3a-06420 & 18~03~01.37 & -23~42~30.8 & 18~03~01.44 & -23~42~29.5 \\
ad3a-06429 & 17~54~05.97 & -23~41~09.1 & 17~54~05.98 & -23~41~08.2 \\
ad3a-06431 & 18~01~02.09 & -23~40~51.6 & 18~01~02.06 & -23~40~49.8 \\
ad3a-06433 & 17~56~42.09 & -23~40~37.2 & 17~56~42.14 & -23~40~35.8 \\
ad3a-06438 & 17~57~37.21 & -23~40~04.7 & 17~57~37.27 & -23~40~03.0 \\
ad3a-06484 & 18~02~00.17 & -23~33~29.7 & 18~02~00.26 & -23~33~28.1 \\
ad3a-06576 & 17~59~35.39 & -23~20~29.4 & 17~59~35.40 & -23~20~27.6 \\
ad3a-06653 & 17~58~58.09 & -23~12~46.9 & 17~58~58.13 & -23~12~44.6 \\
ad3a-06662 & 17~58~05.90 & -23~11~34.3 & 17~58~05.88 & -23~11~32.3 \\
ad3a-06706 & 17~57~34.16 & -23~08~38.3 & 17~57~34.18 & -23~08~37.0 \\
ad3a-25844 & 16~59~16.24 & -40~03~10.2 & 16~59~16.27 & -40~03~10.1 \\
ad3a-25852 & 16~53~26.36 & -40~01~43.6 & 16~53~26.35 & -40~01~44.8 \\
ad3a-25928 & 16~56~43.98 & -39~39~58.6 & 16~56~44.06 & -39~39~59.0 \\
ad3a-25950 & 17~08~32.78 & -39~31~28.9 & 17~08~32.93 & -39~31~28.6 \\
ad3a-25969 & 17~16~56.06 & -39~26~38.8 & 17~16~56.06 & -39~26~38.8 \\
ad3a-26000 & 17~13~21.65 & -39~16~58.9 & 17~13~21.62 & -39~16~58.8 \\
ad3a-26038 & 17~14~28.70 & -39~08~45.5 & 17~14~28.75 & -39~08~45.6 \\
ad3a-26052 & 17~23~09.95 & -39~04~11.3 & 17~23~09.94 & -39~04~11.3 \\
ad3a-26104 & 17~28~01.97 & -38~51~25.9 & 17~28~01.97 & -38~51~26.3 \\
ad3a-26116 & 17~29~32.52 & -38~48~48.6 & 17~29~32.57 & -38~48~47.9 \\
ad3a-26121 & 17~15~18.71 & -38~47~57.4 & 17~15~18.84 & -38~47~57.8 \\
ad3a-26136 & 17~09~46.86 & -38~43~38.2 & 17~09~46.94 & -38~43~37.2 \\
ad3a-26156 & 17~19~48.80 & -38~38~44.9 & 17~19~48.86 & -38~38~45.2 \\
ad3a-26234 & 16~57~07.34 & -38~19~17.3 & 16~57~07.34 & -38~19~17.8 \\
ad3a-26256 & 17~20~02.01 & -38~13~26.0 & 17~20~02.06 & -38~13~25.3 \\
ad3a-26294 & 17~42~17.27 & -38~04~08.2 & 17~42~17.33 & -38~04~08.8 \\
ad3a-26295 & 17~29~41.46 & -38~03~37.7 & 17~29~41.47 & -38~03~39.2 \\
ad3a-26300 & 17~23~28.47 & -38~03~11.5 & 17~23~28.49 & -38~03~11.9 \\
ad3a-26315 & 17~42~11.00 & -37~59~19.2 & 17~42~11.04 & -37~59~20.4 \\
ad3a-26322 & 17~20~34.76 & -37~58~27.2 & 17~20~34.90 & -37~58~27.1 \\
ad3a-26339 & 17~13~21.34 & -37~54~08.9 & 17~13~21.36 & -37~54~08.6 \\
ad3a-26342 & 17~22~10.45 & -37~53~11.6 & 17~22~10.49 & -37~53~11.8 \\
ad3a-26359 & 17~44~39.75 & -37~49~02.8 & 17~44~39.67 & -37~49~04.1 \\
ad3a-26431 & 16~58~25.74 & -37~30~21.2 & 16~58~25.73 & -37~30~21.2 \\
ad3a-26448 & 17~20~40.49 & -37~25~09.0 & 17~20~40.68 & -37~25~09.1 \\
ad3a-26466 & 17~33~33.26 & -37~22~13.1 & 17~33~33.22 & -37~22~13.4 \\
ad3a-26488 & 17~43~44.69 & -37~17~59.3 & 17~43~44.64 & -37~17~59.6 \\
ad3a-26538 & 17~26~59.75 & -37~07~27.5 & 17~26~59.83 & -37~07~27.8 \\
ad3a-26590 & 17~03~32.55 & -36~55~13.7 & 17~03~32.45 & -36~55~14.9 \\
ad3a-26662 & 17~15~38.13 & -36~34~31.1 & 17~15~38.23 & -36~34~32.5 \\
ad3a-26682 & 17~27~51.13 & -36~30~34.8 & 17~27~51.10 & -36~30~34.2 \\
ad3a-26756 & 17~14~09.17 & -36~15~09.9 & 17~14~09.14 & -36~15~10.8 \\
ad3a-26789 & 16~57~52.50 & -36~08~53.7 & 16~57~52.46 & -36~08~53.5 \\
ad3a-26792 & 17~35~15.59 & -36~08~37.0 & 17~35~15.62 & -36~08~37.7 \\
ad3a-26796 & 17~28~12.35 & -36~08~03.6 & 17~28~12.36 & -36~08~02.8 \\
ad3a-26805 & 17~18~39.89 & -36~07~14.2 & 17~18~39.84 & -36~07~13.8 \\
ad3a-26811 & 17~30~16.09 & -36~06~27.2 & 17~30~16.18 & -36~06~27.0 \\
ad3a-26821 & 17~13~49.17 & -36~04~33.3 & 17~13~49.10 & -36~04~32.9 \\
ad3a-26835 & 17~14~28.61 & -36~01~21.1 & 17~14~28.51 & -36~01~21.0 \\
ad3a-26872 & 17~09~19.66 & -35~54~29.5 & 17~09~19.70 & -35~54~29.5 \\
ad3a-26936 & 17~46~28.29 & -35~42~09.9 & 17~46~28.27 & -35~42~09.7 \\
ad3a-26998 & 17~45~24.36 & -35~28~17.0 & 17~45~24.38 & -35~28~17.0 \\
ad3a-27001 & 17~48~39.95 & -35~27~37.5 & 17~48~39.96 & -35~27~38.2 \\
ad3a-27027 & 17~05~05.13 & -35~21~53.2 & 17~05~05.11 & -35~21~52.9 \\
ad3a-27030 & 17~37~07.70 & -35~21~34.4 & 17~37~07.66 & -35~21~34.2 \\
ad3a-27041 & 17~26~42.94 & -35~18~26.4 & 17~26~42.98 & -35~18~24.8 \\
ad3a-27099 & 17~29~05.83 & -35~07~05.6 & 17~29~05.90 & -35~07~05.2 \\
ad3a-27103 & 17~40~32.48 & -35~06~49.7 & 17~40~32.57 & -35~06~50.0 \\
ad3a-27107 & 17~51~53.33 & -35~05~19.9 & 17~51~53.26 & -35~05~19.7 \\
ad3a-27122 & 17~38~25.42 & -35~01~00.5 & 17~38~25.44 & -35~01~00.5 \\
ae3a-00554 & 17~23~15.64 & -36~45~50.2 & 17~23~15.67 & -36~45~49.7 \\
ce3a-00362 & 17~16~16.12 & -37~52~07.4 & 17~16~16.10 & -37~52~05.5 \\
\enddata
\end{deluxetable*}

\end{document}